# Top Quark Mass Measurement from Dilepton Events at CDF II with the Matrix-Element Method

A. Abulencia,[23] D. Acosta,[17] J. Adelman,[13] T. Affolder,[10] T. Akimoto,[55] M.G. Albrow,[16] D. Ambrose,[16] S. Amerio,[43] D. Amidei,[34] A. Anastassov,[52] K. Anikeev,[16] A. Annovi,[18] J. Antos,[1] M. Aoki,[55] G. Apollinari,[16] J.-F. Arguin,[33] T. Arisawa,[57] A. Artikov,[14] W. Ashmanskas,[16] A. Attal,[8] F. Azfar,[42] P. Azzi-Bacchetta,[43] P. Azzurri,[46] N. Bacchetta,[43] H. Bachacou,[28] W. Badgett,[16] A. Barbaro-Galtieri,[28] V.E. Barnes,[48] B.A. Barnett,[24] S. Baroiant,[7] V. Bartsch,[30] G. Bauer,[32] F. Bedeschi,[46] S. Behari,[24] S. Belforte,[54] G. Bellettini,[46] J. Bellinger,[59] A. Belloni,[32] E. Ben Haim,[44] D. Benjamin,[15] A. Beretvas,[16] J. Beringer,[28] T. Berry,[29] A. Bhatti,[50] M. Binkley,[16] D. Bisello,[43] R. E. Blair,[2] C. Blocker,[6] B. Blumenfeld,[24] A. Bocci,[15] A. Bodek,[49] V. Boisvert,[49] G. Bolla,[48] A. Bolshov,[32] D. Bortoletto,[48] J. Boudreau,[47] A. Boveia,[10] B. Brau,[10] C. Bromberg,[35] E. Brubaker,[13] J. Budagov,[14] H.S. Budd,[49] S. Budd,[23] K. Burkett,[16] G. Busetto,[43] P. Bussey,[20] K. L. Byrum,[2] S. Cabrera,[15] M. Campanelli,[19] M. Campbell,[34] F. Canelli,[8] A. Canepa,[48] D. Carlsmith,[59] R. Carosi,[46] S. Carron,[15] M. Casarsa,[54] A. Castro,[5] P. Catastini,[46] D. Cauz,[54] M. Cavalli-Sforza,[3] A. Cerri,[28] L. Cerrito,[42] S.H. Chang,[27] J. Chapman,[34] Y.C. Chen,[1] M. Chertok,[7] G. Chiarelli,[46] G. Chlachidze,[14] F. Chlebana,[16] I. Cho,[27] K. Cho,[27] D. Chokheli,[14] J.P. Chou,[21] P.H. Chu,[23] S.H. Chuang,[59] K. Chung,[12] W.H. Chung,[59] Y.S. Chung,[49] M. Ciljak,[46] C.I. Ciobanu,[23] M.A. Ciocci,[46] A. Clark,[19] D. Clark,[6] M. Coca,[15] G. Compostella,[43] M.E. Convery,[50] J. Conway,[7] B. Cooper,[30] K. Copic,[34] M. Cordelli,[18] G. Cortiana,[43] F. Cresciolo,[46] A. Cruz,[17] C. Cuenca Almenar,[7] J. Cuevas,[11] R. Culbertson,[16] D. Cyr,[59] S. DaRonco,[43] S. D'Auria,[20] M. D'Onofrio,[3] D. Dagenhart,[6] P. de Barbaro,[49] S. De Cecco,[51] A. Deisher,[28] G. De Lentdecker,[49] M. Dell'Orso,[46] F. Delli Paoli,[43] S. Demers,[49] L. Demortier,[50] J. Deng,[15] M. Deninno,[5] D. De Pedis,[51] P.F. Derwent,[16] C. Dionisi,[51] J.R. Dittmann,[4] P. DiTuro,[52] C. Dörr,[25] S. Donati,[46] M. Donega,[19] P. Dong,[8] J. Donini,[43] T. Dorigo,[43] S. Dube,[52] K. Ebina,[57] J. Efron,[39] J. Ehlers,[19] R. Erbacher,[7] D. Errede,[23] S. Errede,[23] R. Eusebi,[16] H.C. Fang,[28] S. Farrington,[29] I. Fedorko,[46] W.T. Fedorko,[13] R.G. Feild,[60] M. Feindt,[25] J.P. Fernandez,[31] R. Field,[17] G. Flanagan,[48] L.R. Flores-Castillo,[47] A. Foland,[21] S. Forrester,[7] G.W. Foster,[16] M. Franklin,[21] J.C. Freeman,[28] I. Furic,[13] M. Gallinaro,[50] J. Galyardt,[12] J.E. Garcia,[46] M. Garcia Sciveres,[28] A.F. Garfinkel,[48] C. Gay,[60] H. Gerberich,[23] D. Gerdes,[34] S. Giagu,[51] P. Giannetti,[46] A. Gibson,[28] K. Gibson,[12] C. Ginsburg,[16] N. Giokaris,[14] K. Giolo,[48] M. Giordani,[54] P. Giromini,[18] M. Giunta,[46] G. Giurgiu,[12] V. Glagolev,[14] D. Glenzinski,[16] M. Gold,[37] N. Goldschmidt,[34] J. Goldstein,[42] G. Gomez,[11] G. Gomez-Ceballos,[11] M. Goncharov,[53] O. González,[31] I. Gorelov,[37] A.T. Goshaw,[15] Y. Gotra,[47] K. Goulianos,[50] A. Gresele,[43] M. Griffiths,[29] S. Grinstein,[21] C. Grosso-Pilcher,[13] R.C. Group,[17] U. Grundler,[23] J. Guimaraes da Costa,[21] Z. Gunay-Unalan,[35] C. Haber,[28] S.R. Hahn,[16] K. Hahn,[45] E. Halkiadakis,[52] A. Hamilton,[33] B.-Y. Han,[49] J.Y. Han,[49] R. Handler,[59] F. Happacher,[18] K. Hara,[55] M. Hare,[56] S. Harper,[42] R.F. Harr,[58] R.M. Harris,[16] K. Hatakeyama,[50] J. Hauser,[8] C. Hays,[15] A. Heijboer,[45] B. Heinemann,[29] J. Heinrich,[45] M. Herndon,[59] D. Hidas,[15] C.S. Hill,[10] D. Hirschbuehl,[25] A. Hocker,[16] A. Holloway,[21] S. Hou,[1] M. Houlden,[29] S.-C. Hsu,[9] B.T. Huffman,[42] R.E. Hughes,[39] J. Huston,[35] J. Incandela,[10] G. Introzzi,[46] M. Iori,[51] Y. Ishizawa,[55] A. Ivanov,[7] B. Iyutin,[32] E. James,[16] D. Jang,[52] B. Jayatilaka,[34] D. Jeans,[51] H. Jensen,[16] E.J. Jeon,[27] S. Jindariani,[17] M. Jones,[48] K.K. Joo,[27] S.Y. Jun,[12] T.R. Junk,[23] T. Kamon,[53] J. Kang,[34] P.E. Karchin,[58] Y. Kato,[41] Y. Kemp,[25] R. Kephart,[16] U. Kerzel,[25] V. Khotilovich,[53] B. Kilminster,[39] D.H. Kim,[27] H.S. Kim,[27] J.E. Kim,[27] M.J. Kim,[12] S.B. Kim,[27] S.H. Kim,[55] Y.K. Kim,[13] L. Kirsch,[6] S. Klimenko,[17] M. Klute,[32] B. Knuteson,[32] B.R. Ko,[15] H. Kobayashi,[55] K. Kondo,[57] D.J. Kong,[27] J. Konigsberg,[17] A. Korytov,[17] A.V. Kotwal,[15] A. Kovalev,[45] A. Kraan,[45] J. Kraus,[23] I. Kravchenko,[32] M. Kreps,[25] J. Kroll,[45] N. Krumnack,[4] M. Kruse,[15] V. Krutelyov,[53] S. E. Kuhlmann,[2] Y. Kusakabe,[57] S. Kwang,[13] A.T. Laasanen,[48] S. Lai,[33] S. Lami,[46] S. Lammel,[16] M. Lancaster,[30] R.L. Lander,[7] K. Lannon,[39] A. Lath,[52] G. Latino,[46] I. Lazzizzera,[43] T. LeCompte,[2] J. Lee,[49] J. Lee,[27] Y.J. Lee,[27] S.W. Lee,[53] R. Lefèvre,[3] N. Leonardo,[32] S. Leone,[46] S. Levy,[13] J.D. Lewis,[16] C. Lin,[60] C.S. Lin,[16] M. Lindgren,[16] E. Lipeles,[9] T.M. Liss,[23] A. Lister,[19] D.O. Litvintsev,[16] T. Liu,[16] N.S. Lockyer,[45] A. Loginov,[36] M. Loreti,[43] P. Loverre,[51] R.-S. Lu,[1] D. Lucchesi,[43] P. Lujan,[28] P. Lukens,[16] G. Lungu,[17] L. Lyons,[42] J. Lys,[28] R. Lysak,[1] E. Lytken,[48] P. Mack,[25] D. MacQueen,[33] R. Madrak,[16] K. Maeshima,[16] T. Maki,[22] P. Maksimovic,[24] S. Malde,[42] G. Manca,[29] F. Margaroli,[5] R. Marginean,[16] C. Marino,[23] A. Martin,[60] V. Martin,[38] M. Martínez,[3] T. Maruyama,[55] P. Mastrandrea,[51] H. Matsunaga,[55] M.E. Mattson,[58] R. Mazini,[33] P. Mazzanti,[5] K.S. McFarland,[49] P. McIntyre,[53] R. McNulty,[29] A. Mehta,[29] S. Menzemer,[11] A. Menzione,[46] P. Merkel,[48] C. Mesropian,[50] A. Messina,[51] M. von der Mey,[8] T. Miao,[16] N. Miladinovic,[6] J. Miles,[32] R. Miller,[35] J.S. Miller,[34] C. Mills,[10] M. Milnik,[25] R. Miquel,[28] A. Mitra,[1] G. Mitselmakher,[17] A. Miyamoto,[26] N. Moggi,[5] B. Mohr,[8] R. Moore,[16] M. Morello,[46] P. Movilla Fernandez,[28] J. Mülmenstädt,[28] A. Mukherjee,[16] Th. Muller,[25] R. Mumford,[24] P. Murat,[16] J. Nachtman,[16] J. Naganoma,[57] S. Nahn,[32] I. Nakano,[40] A. Napier,[56] D. Naumov,[37] V. Necula,[17]

C. Neu,[45] M.S. Neubauer,[9] J. Nielsen,[28] T. Nigmanov,[47] L. Nodulman,[2] O. Norniella,[3] E. Nurse,[30] T. Ogawa,[57] S.H. Oh,[15] Y.D. Oh,[27] T. Okusawa,[41] R. Oldeman,[29] R. Orava,[22] K. Osterberg,[22] C. Pagliarone,[46] E. Palencia,[11] R. Paoletti,[46] V. Papadimitriou,[16] A.A. Paramonov,[13] B. Parks,[39] S. Pashapour,[33] J. Patrick,[16] G. Pauletta,[54] M. Paulini,[12] C. Paus,[32] D.E. Pellett,[7] A. Penzo,[54] T.J. Phillips,[15] G. Piacentino,[46] J. Piedra,[44] L. Pinera,[17] K. Pitts,[23] C. Plager,[8] L. Pondrom,[59] X. Portell,[3] O. Poukhov,[14] N. Pounder,[42] F. Prakoshyn,[14] A. Pronko,[16] J. Proudfoot,[2] F. Ptohos,[18] G. Punzi,[46] J. Pursley,[24] J. Rademacker,[42] A. Rahaman,[47] A. Rakitin,[32] S. Rappoccio,[21] F. Ratnikov,[52] B. Reisert,[16] V. Rekovic,[37] N. van Remortel,[22] P. Renton,[42] M. Rescigno,[51] S. Richter,[25] F. Rimondi,[5] L. Ristori,[46] W.J. Robertson,[15] A. Robson,[20] T. Rodrigo,[11] E. Rogers,[23] S. Rolli,[56] R. Roser,[16] M. Rossi,[54] R. Rossin,[17] C. Rott,[48] A. Ruiz,[11] J. Russ,[12] V. Rusu,[13] H. Saarikko,[22] S. Sabik,[33] A. Safonov,[53] W.K. Sakumoto,[49] G. Salamanna,[51] O. Saltó,[3] D. Saltzberg,[8] C. Sanchez,[3] L. Santi,[54] S. Sarkar,[51] L. Sartori,[46] K. Sato,[55] P. Savard,[33] A. Savoy-Navarro,[44] T. Scheidle,[25] P. Schlabach,[16] E.E. Schmidt,[16] M.P. Schmidt,[60] M. Schmitt,[38] T. Schwarz,[34] L. Scodellaro,[11] A.L. Scott,[10] A. Scribano,[46] F. Scuri,[46] A. Sedov,[48] S. Seidel,[37] Y. Seiya,[41] A. Semenov,[14] L. Sexton-Kennedy,[16] I. Sfiligoi,[18] M.D. Shapiro,[28] T. Shears,[29] P.F. Shepard,[47] D. Sherman,[21] M. Shimojima,[55] M. Shochet,[13] Y. Shon,[59] I. Shreyber,[36] A. Sidoti,[44] P. Sinervo,[33] A. Sisakyan,[14] J. Sjolin,[42] A. Skiba,[25] A.J. Slaughter,[16] K. Sliwa,[56] J.R. Smith,[7] F.D. Snider,[16] R. Snihur,[33] M. Soderberg,[34] A. Soha,[7] S. Somalwar,[52] V. Sorin,[35] J. Spalding,[16] M. Spezziga,[16] F. Spinella,[46] T. Spreitzer,[33] P. Squillacioti,[46] M. Stanitzki,[60] A. Staveris-Polykalas,[46] R. St. Denis,[20] B. Stelzer,[8] O. Stelzer-Chilton,[42] D. Stentz,[38] J. Strologas,[37] D. Stuart,[10] J.S. Suh,[27] A. Sukhanov,[17] K. Sumorok,[32] H. Sun,[56] T. Suzuki,[55] A. Taffard,[23] R. Takashima,[40] Y. Takeuchi,[55] K. Takikawa,[55] M. Tanaka,[2] R. Tanaka,[40] N. Tanimoto,[40] M. Tecchio,[34] P.K. Teng,[1] K. Terashi,[50] S. Tether,[32] J. Thom,[16] A.S. Thompson,[20] E. Thomson,[45] P. Tipton,[49] V. Tiwari,[12] S. Tkaczyk,[16] D. Toback,[53] S. Tokar,[14] K. Tollefson,[35] T. Tomura,[55] D. Tonelli,[46] M. Tönnesmann,[35] S. Torre,[18] D. Torretta,[16] S. Tourneur,[44] W. Trischuk,[33] R. Tsuchiya,[57] S. Tsuno,[40] N. Turini,[46] F. Ukegawa,[55] T. Unverhau,[20] S. Uozumi,[55] D. Usynin,[45] A. Vaiciulis,[49] S. Vallecorsa,[19] A. Varganov,[34] E. Vataga,[37] G. Velev,[16] G. Veramendi,[23] V. Veszpremi,[48] R. Vidal,[16] I. Vila,[11] R. Vilar,[11] T. Vine,[30] I. Vollrath,[33] I. Volobouev,[28] G. Volpi,[46] F. Würthwein,[9] P. Wagner,[53] R. G. Wagner,[2] R.L. Wagner,[16] W. Wagner,[25] R. Wallny,[8] T. Walter,[25] Z. Wan,[52] S.M. Wang,[1] A. Warburton,[33] S. Waschke,[20] D. Waters,[30] W.C. Wester III,[16] B. Whitehouse,[56] D. Whiteson,[45] A.B. Wicklund,[2] E. Wicklund,[16] G. Williams,[33] H.H. Williams,[45] P. Wilson,[16] B.L. Winer,[39] P. Wittich,[16] S. Wolbers,[16] C. Wolfe,[13] T. Wright,[34] X. Wu,[19] S.M. Wynne,[29] A. Yagil,[16] K. Yamamoto,[41] J. Yamaoka,[52] T. Yamashita,[40] C. Yang,[60] U.K. Yang,[13] Y.C. Yang,[27] W.M. Yao,[28] G.P. Yeh,[16] J. Yoh,[16] K. Yorita,[13] T. Yoshida,[41] G.B. Yu,[49] I. Yu,[27] S.S. Yu,[16] J.C. Yun,[16] L. Zanello,[51] A. Zanetti,[54] I. Zaw,[21] F. Zetti,[46] X. Zhang,[23] J. Zhou,[52] and S. Zucchelli[5]

(CDF Collaboration)

[1]*Institute of Physics, Academia Sinica, Taipei, Taiwan 11529, Republic of China*
[2]*Argonne National Laboratory, Argonne, Illinois 60439*
[3]*Institut de Fisica d'Altes Energies, Universitat Autonoma de Barcelona, E-08193, Bellaterra (Barcelona), Spain*
[4]*Baylor University, Waco, Texas 76798*
[5]*Istituto Nazionale di Fisica Nucleare, University of Bologna, I-40127 Bologna, Italy*
[6]*Brandeis University, Waltham, Massachusetts 02254*
[7]*University of California, Davis, Davis, California 95616*
[8]*University of California, Los Angeles, Los Angeles, California 90024*
[9]*University of California, San Diego, La Jolla, California 92093*
[10]*University of California, Santa Barbara, Santa Barbara, California 93106*
[11]*Instituto de Fisica de Cantabria, CSIC-University of Cantabria, 39005 Santander, Spain*
[12]*Carnegie Mellon University, Pittsburgh, PA 15213*
[13]*Enrico Fermi Institute, University of Chicago, Chicago, Illinois 60637*
[14]*Joint Institute for Nuclear Research, RU-141980 Dubna, Russia*
[15]*Duke University, Durham, North Carolina 27708*
[16]*Fermi National Accelerator Laboratory, Batavia, Illinois 60510*
[17]*University of Florida, Gainesville, Florida 32611*
[18]*Laboratori Nazionali di Frascati, Istituto Nazionale di Fisica Nucleare, I-00044 Frascati, Italy*
[19]*University of Geneva, CH-1211 Geneva 4, Switzerland*
[20]*Glasgow University, Glasgow G12 8QQ, United Kingdom*
[21]*Harvard University, Cambridge, Massachusetts 02138*
[22]*Division of High Energy Physics, Department of Physics, University of Helsinki and Helsinki Institute of Physics, FIN-00014, Helsinki, Finland*
[23]*University of Illinois, Urbana, Illinois 61801*
[24]*The Johns Hopkins University, Baltimore, Maryland 21218*
[25]*Institut für Experimentelle Kernphysik, Universität Karlsruhe, 76128 Karlsruhe, Germany*

[26] High Energy Accelerator Research Organization (KEK), Tsukuba, Ibaraki 305, Japan
[27] Center for High Energy Physics: Kyungpook National University, Taegu 702-701, Korea; Seoul National University, Seoul 151-742, Korea; and SungKyunKwan University, Suwon 440-746, Korea
[28] Ernest Orlando Lawrence Berkeley National Laboratory, Berkeley, California 94720
[29] University of Liverpool, Liverpool L69 7ZE, United Kingdom
[30] University College London, London WC1E 6BT, United Kingdom
[31] Centro de Investigaciones Energeticas Medioambientales y Tecnologicas, E-28040 Madrid, Spain
[32] Massachusetts Institute of Technology, Cambridge, Massachusetts 02139
[33] Institute of Particle Physics: McGill University, Montréal, Canada H3A 2T8; and University of Toronto, Toronto, Canada M5S 1A7
[34] University of Michigan, Ann Arbor, Michigan 48109
[35] Michigan State University, East Lansing, Michigan 48824
[36] Institution for Theoretical and Experimental Physics, ITEP, Moscow 117259, Russia
[37] University of New Mexico, Albuquerque, New Mexico 87131
[38] Northwestern University, Evanston, Illinois 60208
[39] The Ohio State University, Columbus, Ohio 43210
[40] Okayama University, Okayama 700-8530, Japan
[41] Osaka City University, Osaka 588, Japan
[42] University of Oxford, Oxford OX1 3RH, United Kingdom
[43] University of Padova, Istituto Nazionale di Fisica Nucleare, Sezione di Padova-Trento, I-35131 Padova, Italy
[44] LPNHE, Universite Pierre et Marie Curie/IN2P3-CNRS, UMR7585, Paris, F-75252 France
[45] University of Pennsylvania, Philadelphia, Pennsylvania 19104
[46] Istituto Nazionale di Fisica Nucleare Pisa, Universities of Pisa, Siena and Scuola Normale Superiore, I-56127 Pisa, Italy
[47] University of Pittsburgh, Pittsburgh, Pennsylvania 15260
[48] Purdue University, West Lafayette, Indiana 47907
[49] University of Rochester, Rochester, New York 14627
[50] The Rockefeller University, New York, New York 10021
[51] Istituto Nazionale di Fisica Nucleare, Sezione di Roma 1, University of Rome "La Sapienza," I-00185 Roma, Italy
[52] Rutgers University, Piscataway, New Jersey 08855
[53] Texas A&M University, College Station, Texas 77843
[54] Istituto Nazionale di Fisica Nucleare, University of Trieste/ Udine, Italy
[55] University of Tsukuba, Tsukuba, Ibaraki 305, Japan
[56] Tufts University, Medford, Massachusetts 02155
[57] Waseda University, Tokyo 169, Japan
[58] Wayne State University, Detroit, Michigan 48201
[59] University of Wisconsin, Madison, Wisconsin 53706
[60] Yale University, New Haven, Connecticut 06520

We describe a measurement of the top quark mass using events with two charged leptons collected by the CDF II detector from $p\overline{p}$ collisions with $\sqrt{s} = 1.96$ TeV at the Fermilab Tevatron. The likelihood in top quark mass is calculated for each event by convoluting the leading order matrix element describing $q\overline{q} \to t\overline{t} \to b\ell\nu_\ell\overline{b}\ell'\nu_{\ell'}$ with detector resolution functions. The presence of background events in the data sample is modeled using similar calculations involving the matrix elements for major background processes. In a data sample with integrated luminosity of 340 pb$^{-1}$, we observe 33 candidate events and measure $M_{top} = 165.2 \pm 6.1(\text{stat.}) \pm 3.4(\text{syst.})$ GeV/$c^2$. This measurement represents the first application of this method to events with two charged leptons and is the most precise single measurement of the top quark mass in this channel.

## I. INTRODUCTION

The standard model accommodates quark masses through Yukawa couplings to the Higgs boson, but does not predict the size of these couplings and contains no explanation for the observed quark masses. A striking feature is the large mass of the top quark, the heaviest of the observed fundamental particles. Its large mass suggests that it may play a unique role in electroweak symmetry breaking [1, 2]. Precise measurements of the mass of the top quark constrain the mass of the yet unobserved Higgs boson through radiative corrections [3], and can restrict possible extensions to the standard model [4].

In collisions of protons and anti-protons at $\sqrt{s} = 1.96$ TeV, top quark pairs are produced primarily through the annihilation of quarks and anti-quarks. In the standard model, the top quark decays to a $b$ quark and a $W$ boson nearly 100% of the time; the $W$ boson decays to a pair of quarks or a charged lepton and neutrino. Quarks fragment and hadronize and are reconstructed as jets (clusters of particles). The dilepton channel, consisting of decays $t\bar{t} \to b\ell\nu_\ell \bar{b}\ell'\nu_{\ell'}$, has a small branching fraction, but measurements of the mass in this channel have the advantage that they are less reliant on the calibration of the jet energy scale than channels with hadronic $W$ boson decay. A top quark mass measurement in this channel is an important verification that the observed top quark candidates are consistent with standard model production and decay. A discrepancy from measurements in other channels could indicate the presence of physics beyond the standard model that makes contributions to the dilepton sample [5].

Reconstruction of the top quark mass in the dilepton channel is particularly challenging due to the two undetected neutrinos from the $W$ boson decays. Previous measurements [6, 7] in this channel using Tevatron run I data calculated a mass in each event by making several kinematic assumptions and integrating over the remaining unmeasured quantities; the distribution of event masses was then compared to simulation at varying top quark masses. This article describes in detail the first application to the dilepton channel of a technique pioneered for analysis of single lepton $t\bar{t} \to b\ell\nu_\ell \bar{b}qq'$ decays [8, 9, 10, 11, 12]. This technique convolutes the matrix element for $t\bar{t}$ decays with detector resolution functions and integrates over unmeasured quantities to construct per-event likelihoods in top quark mass. We relax many of the kinematic assumptions of previous methods and integrate over six unmeasured quantities. The event likelihoods are directly multiplied to obtain the joint likelihood from which $M_t$ is extracted. This weights events according to the relative amount of information they carry. The data used in this measurement correspond to an integrated luminosity of 340 pb$^{-1}$ collected between March 2002 and August 2004 by the CDF II detector in collisions at $\sqrt{s} = 1.96$ TeV at the Fermilab Tevatron; this measurement and a brief description were first reported in Ref. [13].

Sections II and III describe the CDF detector and the selection of the data sample. Section IV gives an overview of the analysis method. Sections V, VI and VII describe in detail the major pieces of the likelihood calculation. Section VIII describes the reconstruction and calibration of the top quark mass. Section IX covers the systematic uncertainties and Section X presents the measurement.

## II. THE CDF DETECTOR

The CDF II detector is an azimuthally and forward-backward symmetric detector designed to study $p\bar{p}$ collisions at the Fermilab Tevatron. The CDF coordinate system is right-handed, with the $z$-axis pointing along a tangent to the Tevatron ring along the proton direction. The remaining rectangular coordinates $x$ and $y$ are defined pointing outward and upward from the Tevatron ring respectively, and the azimuthal angle $\phi$ is measured relative to the $x$ axis in the $xy$-plane. Transverse quantities such as transverse momentum, $p_T$, and transverse energy, $E_T$, are projections onto this plane. The polar angle $\theta$ is measured from the proton direction and is typically expressed as pseudorapidity $\eta = -\ln(\tan\frac{\theta}{2})$. Subdetectors which are particularly relevant to this analysis are described below. A more complete description of the CDF II detector can be found elsewhere [14].

The CDF tracking system consists of an inner silicon microstrip detector and a large outer open-cell drift chamber. These subsystems are immersed in a superconducting solenoid producing a 1.4 T magnetic field parallel to the $p$ and $\bar{p}$ beams. The silicon detector, which provides high-resolution position measurements of charged particles close to the interaction region, consists of three subdetectors. The innermost detector, Layer 00 (L00) [15], is a single-sided layer of silicon wafers mounted directly on the beampipe at a radius of 1.6 cm. The SVXII [16] detector is a five layer, double-sided silicon detector that covers the radial region between 2.5 cm and 10.6 cm. The Intermediate Silicon Layers (ISL) [17] comprise one or two additional layers of double-sided silicon, depending on the polar angle, at radii from 20 cm to 28 cm. The Central Outer Tracker (COT) [18], a large open-cell drift chamber, is positioned outside the silicon detector from radii of 0.43 m to 1.32 m. The COT contains 8 superlayers (alternating between axial and $\pm 2°$ stereo angle) each containing 12 wire layers for a total of 96 layers. In combination the Silicon and COT detectors provide excellent tracking up to $|\eta| \leq 1.1$ with decreasing coverage to $|\eta| \leq 2.0$.

Sampling calorimeters segmented in $\eta$ and $\phi$ surrounding the tracking system measure particle energies. In the central region of $|\eta| < 1.1$, the calorimeter is divided into projective towers subtending 15° in $\phi$ and 0.1 in $\eta$. The central electromagnetic calorimeter (CEM) [19] constitutes the front of the wedges in the central region. The CEM consists of alternating layers of lead and scintillator, amounting to 18 radiation lengths of material. Embedded in the CEM is the shower maximum detector (CES). The CES provides position measurements of the electromagnetic showers at a depth of 5 radiation lengths and is used in electron identification. Behind the CEM is the central hadronic calorimeter (CHA) [20], which provides energy measurements of hadronic jets. The CHA consists of 4.7 interaction lengths of alternating steel and scintillator. In addition to the central calorimeters, end plug calorimeters cover $1.1 < |\eta| < 3.6$. The plug electromagnetic calorimeter (PEM) [21] consists of alternating lead absorber and scintillating tile readout with wavelength shifting fibers; the total thickness is 23.2 radiation lengths of material. A plug shower maximum detector (PES) [22] provides position measurement of electron and photon showers. The plug hadronic calorimeter (PHA) has alternating layers of iron and scintillating tile for a total of 6.8 interaction lengths.

The muon detection system consists of three sandwiched drift tube layers, each utilizing single wire drift cells four layers deep. Directly behind the central hadronic calorimeter is the central muon detector (CMU) [23] which can detect muons with $p_T > 1.4$ GeV/$c$ in the region of $|\eta| < 0.6$. Additional muon coverage in this region is provided by the central muon upgrade (CMP) which is separated from the CMU by 60 cm of steel. The CMP detects muons with $p_T > 2.0$ GeV/$c$. The central muon extension (CMX) provides further coverage in the region of $0.6 < |\eta| < 1.0$.

CDF's three level trigger system reduces the event rate from 1.7 MHz to $\approx 80$ Hz. The first two levels are hardware triggers that partially reconstruct events using information from individual subdetectors while the third level is a software trigger that performs event reconstruction.

## III. DATA SAMPLE

We select $t\bar{t} \to b\ell\nu_\ell \bar{b}\ell'\overline{\nu}_{\ell'}$ decays with a high-$p_T$ lepton trigger and the requirement that candidates have (i) two leptons each with $p_T > 20$ GeV/$c$, (ii) significant missing energy transverse to the beam direction ($\not\!\!E_T$) [24], and (iii) two jets each with $E_T > 15$ GeV. The selection was designed for a cross-section measurement and is described as "DIL" in Ref. [25]. A description of the trigger requirements and selection used to obtain this dataset follows.

### A. Trigger

The trigger requires at least one high-$p_T$ lepton. For central electron candidates, the first two trigger levels require an electromagnetic calorimeter cluster with a confirming COT track and without a large hadronic energy deposit. The third level trigger requires an electron candidate with $E_T \geq 18$ GeV. Events with electron candidates in the plug ($|\eta| > 1.2$) are required to have electron $E_T > 20$ GeV and missing transverse energy $\not\!\!E_T > 15$ GeV. For muon candidates, the first two trigger levels require hits in the muon chambers and a confirming COT track. The third level trigger requires a muon stub with a matching track of $p_T \geq 18$ GeV/$c$.

### B. Leptons

Final lepton requirements are tighter than those made in the last stage of the trigger. Electron candidates are required to have an electromagnetic calorimeter cluster with $E_T > 20$ GeV and $|\eta| < 2.0$. Muon candidates are required to have a track with $p_T > 20$ GeV/$c$, which limits them to the $\eta$ coverage of the COT. At least one of the leptons is required to be isolated. Electrons and muons are isolated if the energy deposited in the calorimeter in a cone $\Delta R \equiv \sqrt{(\Delta\eta)2 + (\Delta\phi)2} = 0.4$ around the lepton is at most 10% of the lepton transverse energy. In addition, electron candidates are required to have a well-measured track pointing at an energy deposition in the calorimeter. For electron candidates with $|\eta| > 1.2$, this track association uses a calorimeter-seeded silicon tracking algorithm [26]. Muon candidates are required to have a well-measured track linked to hits in the muon chambers and energy deposition in the calorimeter consistent with that expected for muons. If the event contains two muons, only one is required to have hits in muon chambers used in the trigger decision. The other muon may have hits in chambers not used for the trigger decision if there is a matching COT track, or no hits in muon chambers if the COT track points in regions where there is no muon chamber coverage.

### C. Jets

A jet is defined as a cluster of energy surrounding a calorimeter tower with $E_T > 1$ GeV, grouped within a cone of $\Delta R \equiv \sqrt{(\Delta\eta)^2 + (\Delta\phi)^2} = 0.4$ by the JETCLU algorithm [27]. Events are required to have at least two jets with $|\eta| < 2.5$ and $E_T > 15$ GeV after the corrections described below are applied.

The total jet $E_T$ is corrected for non-uniformities in the response of the calorimeter as a function of $\eta$, for effects from multiple $p\overline{p}$ collisions, and for the hadronic jet energy scale of the calorimeter [28]. The two highest $E_T$ jets for each event are assumed to stem from the $b$ quarks; this assumption is true for $\sim$70% of simulated $t\bar{t}$ events. The momentum components of each $b$ quark are then calculated from the measured jet $E_T$ and angle by assuming a $b$ quark mass of 4.7 GeV/$c^2$ [29]. No explicit identification of $b$ jets is used.

### D. Final Selection Cuts

After lepton and jet identification, further requirements are made to reduce the expected level of background in the sample. Events are required to have missing transverse energy of $\not\!\!E_T > 25$ GeV. $\not\!\!E_T$ is corrected for the presence of isolated high-$p_T$ muons by subtracting the momentum lost by the muons in the calorimeter and adding the muon $p_T$ to the vector sum. In events with $\not\!\!E_T < 50$ GeV, the direction of the $\not\!\!E_T$ vector is required to be separated by at least 20° in $\phi$ from any lepton or jet in the event. This reduces the background from Drell-Yan production of $\tau$ pairs as well as the number of events in which mismeasured jet or lepton energy contributes a large fraction of the $\not\!\!E_T$.

To reduce the number of $Z/\gamma^* \to ee, \mu\mu$ events in which mismeasured jet energy leads to significant amounts of measured missing tranverse energy, $ee$ and $\mu\mu$ events with dilepton invariant mass between 76 and 106 GeV/$c^2$ are required to have their $\not\!\!E_T$ vector point away from any energetic jets in the event [30].

To further suppress background, events are required to have $H_T$, defined as the scalar sum of lepton $E_T$, the $E_T$ of the two leading jets, and $\not\!\!E_T$, greater than 200 GeV.

Events which are likely to be due to cosmic rays are removed by requiring a coincidence of the muon arrival times to the calorimeter. Electrons resulting from photon conversion to $e^+e^-$ pairs are also removed. Conversions are identified by pairing the electron track to a track of opposite sign and requiring that the two tracks are consistent with originating from a common vertex and being parallel at that vertex. Events with three leptons are removed as well as events in which the leptons have the same sign.

### E. Sample Composition

Table I lists the number of expected background events of each type and the number of $t\bar{t}$ signal events expected at various top quark masses [31] for the data sample used in this measurement. The signal estimate includes $t\bar{t}$ events in which a $W$ boson decays to a $\tau$ when the $\tau$ decays to an $e$ or a $\mu$. Studies in Monte Carlo simulations show that 14% of the accepted signal events have at least one $W$ boson decaying to a $\tau$.

The largest source of expected background is $Z/\gamma^* \to ee, \mu\mu$ events with associated jets. The expected contribution to the sample from this background is estimated using a combination of $Z$ boson data and PYTHIA [32] Monte Carlo events. The second largest source of expected backgrounds is events in which one or more jets are misidentified as leptons. The probability for a jet to be misidentified as a lepton is very small, so the majority of these events have one real lepton and one jet misidentified as a lepton. The expected contribution to the sample from this background is estimated using $W \to \ell\nu$+jets data. There are several smaller sources of expected background: $WW, WZ$ and $Z/\gamma^* \to \tau\tau$ with leptonic $\tau$ decays. The expected contribution from these processes is estimated using ALPGEN [33] and PYTHIA Monte Carlo events. Other processes make negligible contributions. A detailed description of the method of background estimation used can be found in Ref. [25].

TABLE I: Expected numbers of signal and background events for a data sample of $\int \mathcal{L}dt = 340\text{pb}^{-1}$. The signal cross section is obtained from [31]. The total expected background is the sum of the indented background contributions.

| Source | | Events |
|---|---|---|
| $t\bar{t}$ ($M_t = 165$ GeV/$c^2$, $\sigma = 9.1$pb) | | $21.7 \pm 1.4$ |
| $t\bar{t}$ ($M_t = 175$ GeV/$c^2$, $\sigma = 6.7$pb) | | $17.2 \pm 1.4$ |
| $t\bar{t}$ ($M_t = 185$ GeV/$c^2$, $\sigma = 4.9$pb) | | $13.3 \pm 1.4$ |
| Total Expected Background Rate | | $10.5 \pm 1.9$ |
| $WW$ | $1.2 \pm 0.2$ | |
| $WZ$ | $0.4 \pm 0.1$ | |
| $Z/\gamma^* \to ee, \mu\mu$ | $4.7 \pm 1.2$ | |
| $Z/\gamma^* \to \tau\tau$ | $0.8 \pm 0.2$ | |
| Lepton misID | $3.5 \pm 1.4$ | |
| Total Expected Rate ($M_t = 165$ GeV/$c^2$) | | $32.2 \pm 2.3$ |
| Total Expected Rate ($M_t = 175$ GeV/$c^2$) | | $27.7 \pm 2.3$ |
| Total Expected Rate ($M_t = 185$ GeV/$c^2$) | | $23.8 \pm 2.3$ |
| Run II Data ($\int \mathcal{L}dt = 340\text{pb}^{-1}$) | | 33 |

## IV. ANALYSIS OVERVIEW

The probability density for $t\bar{t}$ decays is expressed as $P_s(\mathbf{x}|M_t)$, where $M_t$ is the top quark pole mass and $\mathbf{x}$ contains the lepton and jet momentum measurements. We calculate $P_s(\mathbf{x}|M_t)$ using the theoretical description of the $t\bar{t}$ production and decay process expressed with respect to $\mathbf{x}$,

$$P_s(\mathbf{x}|M_t) = \frac{1}{\sigma(M_t)} \frac{d\sigma(M_t)}{d\mathbf{x}}, \tag{1}$$

where $\frac{d\sigma}{d\mathbf{x}}$ is the differential cross section evaluated with respect to event measurements contained in $\mathbf{x}$.

To evaluate the differential cross section $\frac{d\sigma}{d\mathbf{x}}$, we convolute the leading order matrix-element $\mathcal{M}$ for $q\bar{q} \to t\bar{t} \to bl\nu_l\bar{b}l'\nu_{l'}$ with detector resolution functions and integrate over unmeasured quantities.

The matrix element depends on the momenta of the incoming partons ($q_1$ and $q_2$), and of the outgoing two $b$ quarks ($p_1$ and $p_2$), two leptons ($\ell_1$ and $\ell_2$), and two neutrinos ($\nu_1$ and $\nu_2$). Observed quantities consist of jets $j_1$ and $j_2$, measured leptons $L_1$ and $L_2$, and two components of missing transverse energy. To express the differential cross section with respect to these observed quantities $\mathbf{x}$, transfer functions are introduced to connect the quantities which correspond to external legs of the matrix element $(q_1, q_2, p_1, p_2, l_1, l_2, \nu_1, \nu_2)$ to the observed quantities $(j_1, j_2, L_1, L_2)$. Quantities which are well-measured by the detector, lepton momenta and jet angles, are described by delta functions which directly reduce the number of unknown parton-level quantities. Integrations are performed over quantities which are not directly measured, *i.e.* quark and neutrino energies. While quark energies are not directly measured, they can be estimated from the observed energies of the corresponding jets. The transfer function between quark and jet energies parameterizes this relationship, and is expressed as $W(E_p, E_j)$, the probability of measuring jet energy $E_j$ given parton energy $E_p$.

We make the following assumptions regarding the transfer between parton-level quantities and the observables:

- Leptons are measured perfectly. We express the lepton transfer functions as a three-dimensional $\delta$-function,

$$\delta^3(\ell_1 - L_1)\, \delta^3(\ell_2 - L_2).$$

- Jet angles are measured perfectly, and jet energy

can be described as a parametric function of parton energy. We express the $b$ quark to jet transfer function as

$$\delta(\theta_{j_1} - \theta_{p_1})\delta(\phi_{j_1} - \phi_{p_1})W(E_{p_1}, E_{j_1}).$$

- The two most energetic jets in the event are due to the fragmentation and hadronization of the two $b$ quarks from top quark decay. All other jets in the event, if present, are disregarded.
- Incoming partons are massless and have no transverse momentum.
- Masses of the final state leptons are zero, masses of the $b$ quarks are set to 4.7 GeV/$c^2$.

The probability density in $\mathbf{x}$ for $q\overline{q} \to t\overline{t} \to \overline{b}\ell^-\overline{\nu}_\ell b\ell'^+\nu'_\ell$ for a fixed $M_t$ can be written as

$$P_s(\mathbf{x}|M_t) = \frac{1}{\sigma(M_t)} \int d\Phi \sum_{a,b} |\mathcal{M}_{t\bar{t}}(q_i, p_i; M_t)|^2 \times \prod_{jets} W(E_p, E_j) f^a_{\mathtt{PDF}}(q_1) f^b_{\mathtt{PDF}}(q_2). \quad (2)$$

In this expression, the integral is over the phase space $d\Phi$ for $q\overline{q} \to t\overline{t} \to bl\nu_l\overline{b}l'\nu_{l'}$, the sum runs over the flavors $a, b$ of the incoming partons, and $f^a_{\mathtt{PDF}}$ are parton distribution functions for flavor $a$. Constraints such as conservation of momentum which appear as delta functions and modify the integration are here implicitly included in the phase-space integration and are discussed in detail in the following sections. The term, $1/\sigma(M_t)$, in front of the integral ensures the normalization condition for the probability,

$$\int d\mathbf{x}\ P_s(\mathbf{x}|M_t) = 1, \quad (3)$$

where the integration is performed over all accepted $\mathbf{x}$ to account for mass-dependent effects of the selection.

### A. Signal and Background Processes

We calculate the probability for the dominant background processes, $P_{bg}(\mathbf{x})$ and form the generalized per-event probability density in $\mathbf{x}$,

$$P(\mathbf{x}|M_t) = P_s(\mathbf{x}|M_t)p_s(M_t) + P_{bg_1}(\mathbf{x})p_{bg_1} + P_{bg_2}(\mathbf{x})p_{bg_2} + \cdots, \quad (4)$$

as a weighted sum of the probabilities for each process, where the weights $p_s(M_t)$ and $p_{bg_i}$ are determined from the expected fractions of signal and background events (see Table I). We evaluate probabilities for the three largest expected backgrounds: $Z/\gamma^*$ boson production with 2 associated jets, $W$ boson production with 3 associated jets in which one jet is incorrectly identified as a lepton, and $W$ boson pair production with 2 associated jets.

### B. Calibration

The probability $P(\mathbf{x})$ is an approximation of the true probability and balances precision with computational tractability. To account for the approximations made in the construction of $P(\mathbf{x})$, we derive a calibration in fully realistic Monte Carlo events. This strategy accounts for the differences between the model used in constructing $P(\mathbf{x})$ and the fully realistic simulation. Uncertainties in the model used to produce the simulated events contribute to systematic errors on the final measurement described in Section VIII.

## V. TRANSFER FUNCTIONS

The transfer function between quark and jet energies, $W(E_p, E_j)$, expresses the probability of measuring jet energy $E_j$ from a given parton with energy $E_p$ such that

$$n(E_j, E_p)dE_j dE_p = n(E_p)dE_p W(E_p, E_j), \quad (5)$$

where $n(E_j, E_p)dE_j dE_p$ is the number of events with jet energy between $E_j$ and $E_j + dE_j$ and parton energy between $E_p$ and $E_p + dE_p$, and $n(E_p)$ is the number of partons with energy between $E_p$ and $E_p + dE_p$.

We parameterize the distribution of measured jet energies, $E_j$, as a function of the quark energies, $E_p$, and the difference between the parton energy and the jet energy, $W(\delta \equiv E_p - E_j)$ [12]. The parameterization is a sum of two Gaussians to account for both the peak of the $\delta$ distribution and its tails,

$$W(\delta) = \frac{1}{\sqrt{2\pi}(p_2 + p_3 p_5)} \left[ e^{\frac{-(\delta - p_1)^2}{2p_2^2}} + p_3 e^{\frac{-(\delta - p_4)^2}{2p_5^2}} \right], \quad (6)$$

where each $p_i$ depends linearly on $E_p$:

$$p_i = a_i + b_i E_p. \quad (7)$$

The parameters $p_i$ are extracted with an unbinned maximum likelihood fit over $N$ jets in a sample of simulated HERWIG [34] $t\bar{t}$ events with $M_t = 178$ GeV/$c^2$ which pass the event selection and contain jets whose axis is contained in a cone of $\Delta R = 0.4$ surrounding $b$ quarks. Jets which arise from initial or final state radiation are excluded. The log likelihood is expressed as a sum over jets:

$$-\ln L = -\sum_{k=1}^{N} \ln n(E_{p_k}) - \sum_{k=1}^{N} \ln W(E_{p_k}, E_{j_k}). \quad (8)$$

The first term does not depend on the parameters $p_i$ and can be dropped from the minimization. We extract the parameters shown in Table II.

TABLE II: Parameters for $W(E_p, E_j)$ extracted using jets matched in angle to $b$ quarks (see text), from HERWIG $t\bar{t}$ Monte Carlo.

| $p_i$ | $a_i$ | $b_i$ |
|---|---|---|
| $p_1$ | $1.90 \pm 0.62$ GeV | $0.023 \pm 0.008$ |
| $p_2$ | $2.83 \pm 0.54$ GeV | $0.075 \pm 0.005$ |
| $p_3$ | $0.70 \pm 0.08$ | $0.000 \pm 0.001$ GeV$^{-1}$ |
| $p_4$ | $-1.79 \pm 0.79$ GeV | $-0.187 \pm 0.012$ |
| $p_5$ | $8.04 \pm 0.67$ GeV | $0.095 \pm 0.008$ |

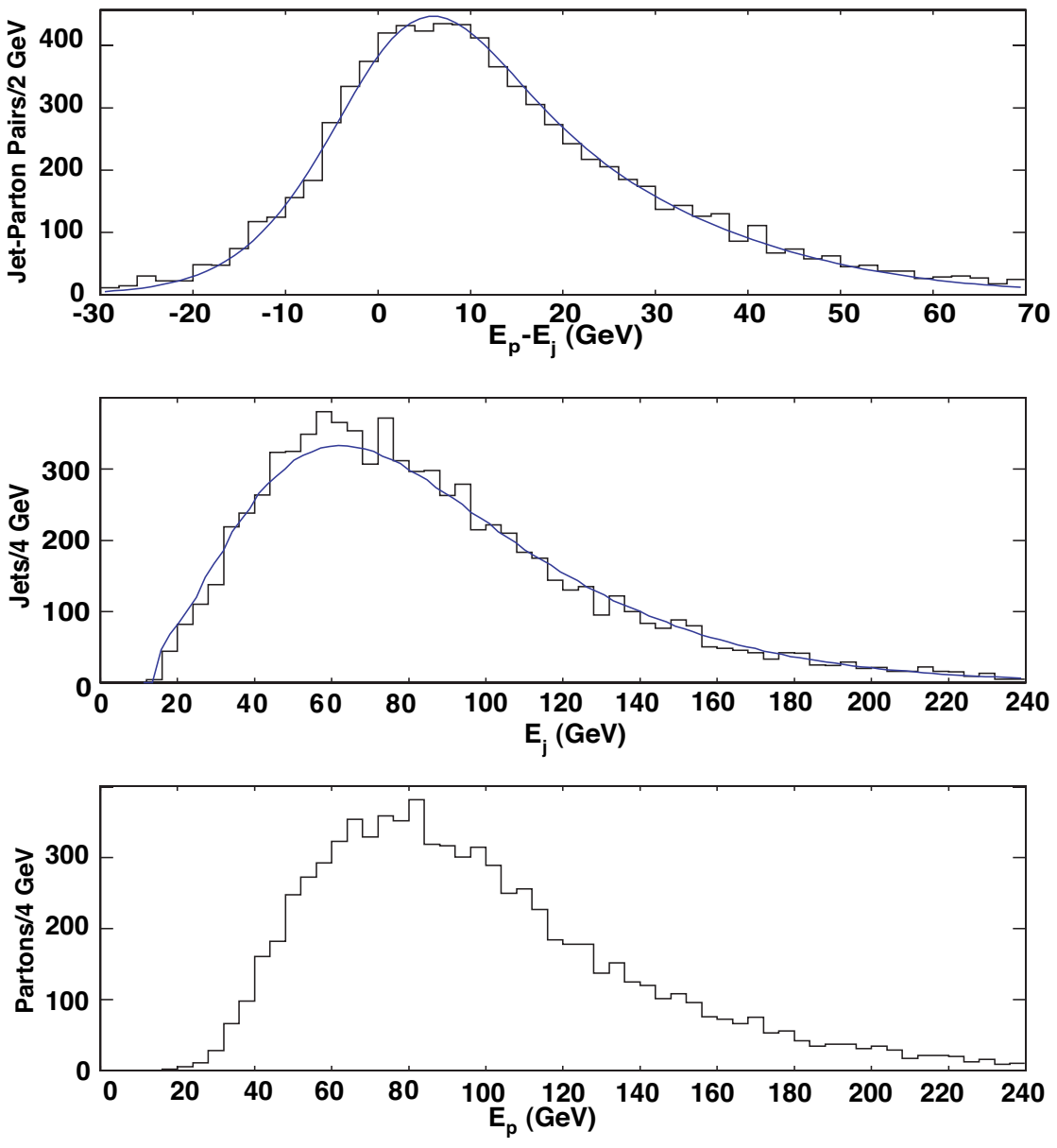

FIG. 1: Top, difference of $E_p - E_j$ between parton and jet energy for reconstructed jets matched in angle to $b$ quarks (see text). Center, distribution of $E_j$ of jet energy. Bottom, input distribution of parton energy $E_p$. Histograms are simulated events for $M_t = 178$ GeV/$c^2$; curves in the upper two histograms show the distributions calculated using $W(E_p, E_j)$.

To test the jet transfer function, we calculate the distribution of jet energies which result from simulated partons of known energy. The calculation for jet energies resulting from partons with energy $E_1 < E_p < E_2$ is the integral of $n(E_j, E_p)dE_p$:

$$\int_{E_1}^{E_2} n(E_p) dE_p W(E_p, E_j). \tag{9}$$

The calculation for the jet energy distribution and the difference in jet and parton energies resulting from all partons ( $0 < E_p < 1$ TeV ) in a simulated sample of $t\bar{t}$ with $M_t = 178$ GeV/$c^2$ is shown in Figure 1. Similar tests using slices of parton energies are shown in Figure 2.

The jet transfer function models the detector response to partons and should be independent of the production process. We confirm this by using $W(E_p, E_j)$ parameterized from $M_t = 178$ GeV/$c^2$ events to calculate the jet energy distribution resulting from $b$ quarks in Monte Carlo top quark decays of varied top quark masses. Figure 3 shows that the jet transfer function derived using partons from $M_t = 178$ GeV/$c^2$ top quark decays satisfactorily describes jet energies from top quark decays of $M_t$ ranging from 150 GeV/$c^2$ through 200 GeV/$c^2$. The performance of the transfer functions in fully realistic simulated events is included in the measurement calibrations discussed below.

## VI. SIGNAL LIKELIHOOD

The probability density for $q\bar{q} \to t\bar{t} \to b\ell\nu_\ell\bar{b}\ell'\nu_{\ell'}$ decays is constructed as the differential cross section, $d\sigma$, with respect to the measured event quantities, $\mathbf{x}$. The total cross section $\sigma$ is written as

$$\sigma = \int \sum_{a,b} \frac{(2\pi)^4 |\mathcal{M}|^2}{4\sqrt{(q_1 \cdot q_2)^2 - m_1^2 m_2^2}} \times f^a_{\text{PDF}}\left(\frac{q_{z_1}}{E_{\text{beam}}}\right) f^b_{\text{PDF}}\left(\frac{q_{z_2}}{E_{\text{beam}}}\right) d\Phi_6 dq_1 dq_2, \tag{10}$$

where the sum runs over incoming parton flavors, $\mathcal{M}$ is the matrix element for the process, $q_{1,2}$ and $m_{1,2}$ refer to the momenta and mass of the incoming partons, $f_{PDF}$ are the parton distribution functions for flavor $a$, and the integration is over the phase space for the six final state particles as well as the longitudinal momenta of the incoming particles.

The matrix element [35, 36] has the form

$$|\mathcal{M}|^2 = \frac{g_s^4}{9} F\bar{F}\left((2 - \beta^2 s_{qt}^2) - X_{sc}\right), \tag{11}$$

where $\beta$ is the top quark velocity in the $q\bar{q}$ rest frame, $X_{sc}$ contains terms describing spin correlations between the top quarks, $g_s$ is the strong coupling constant ($g_s^2/4\pi = \alpha_s$), $s_{qt}$ is the sine of the angle between the incoming parton and the top quark, and $F$ and $\overline{F}$ are the propagators for the top and the anti-top quark respectively. We drop the spin correlation term $X_{sc}$ as it is negligible. The top quark propagator and decay terms are given by

$$F = \frac{g_w^4}{4}\left[\frac{m_t^2 - m_{\bar{\ell}\nu}^2}{(m_t^2 - M_t^2)^2 + (M_t\Gamma_t)^2}\right] \times \left[\frac{m_t^2(1 - \hat{c}_{\bar{\ell}b}^2) + m_{\bar{\ell}\nu}^2(1 + \hat{c}_{\bar{\ell}b})^2}{(m_{\bar{\ell}\nu}^2 - M_W^2)^2 + (M_W\Gamma_W)^2}\right], \tag{12}$$

where $m_t$ is the invariant mass of the $t$ quark decay products and $\hat{c}_{ij}$ is the cosine of the angle between particles $i$ and $j$ in the $W$ boson rest frame. The $M_t$, $\Gamma_t$, $M_W$, $\Gamma_W$ are the pole masses and widths of the top quark and $W$ boson, and $g_w$ is the weak coupling constant. The top quark width, $\Gamma_t$, is a function of $M_t, M_W$ [12] and $\Gamma_W$ as described by the standard model. $\overline{F}$ is given by the same expression as Eq. 12, replacing the terms for $t$ and its decay products with $\bar{t}$ and its decay products.

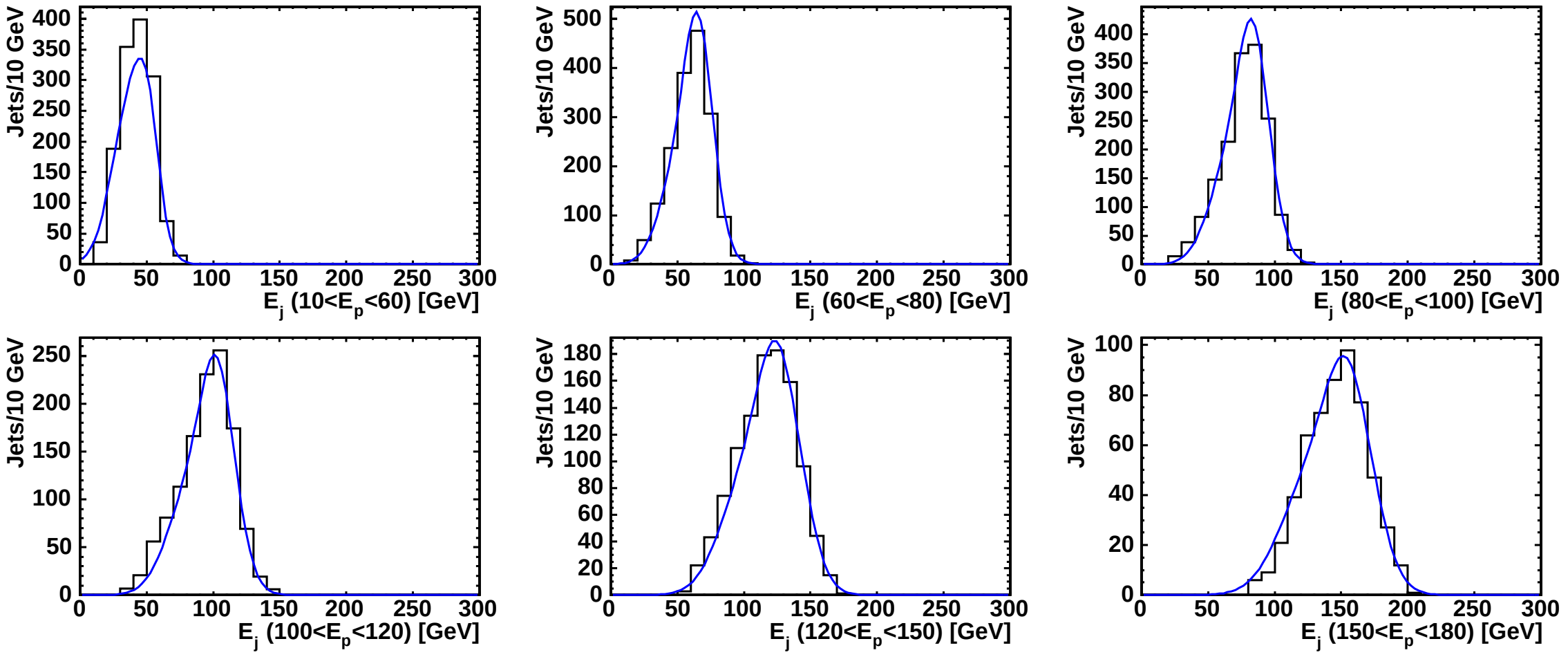

FIG. 2: Comparison of simulated $E_j$ with calculations from $W(E_p, E_j)$ from six ranges of $E_p$. Histograms are simulated events; curves show the calculated distributions using $W(E_p, E_j)$.

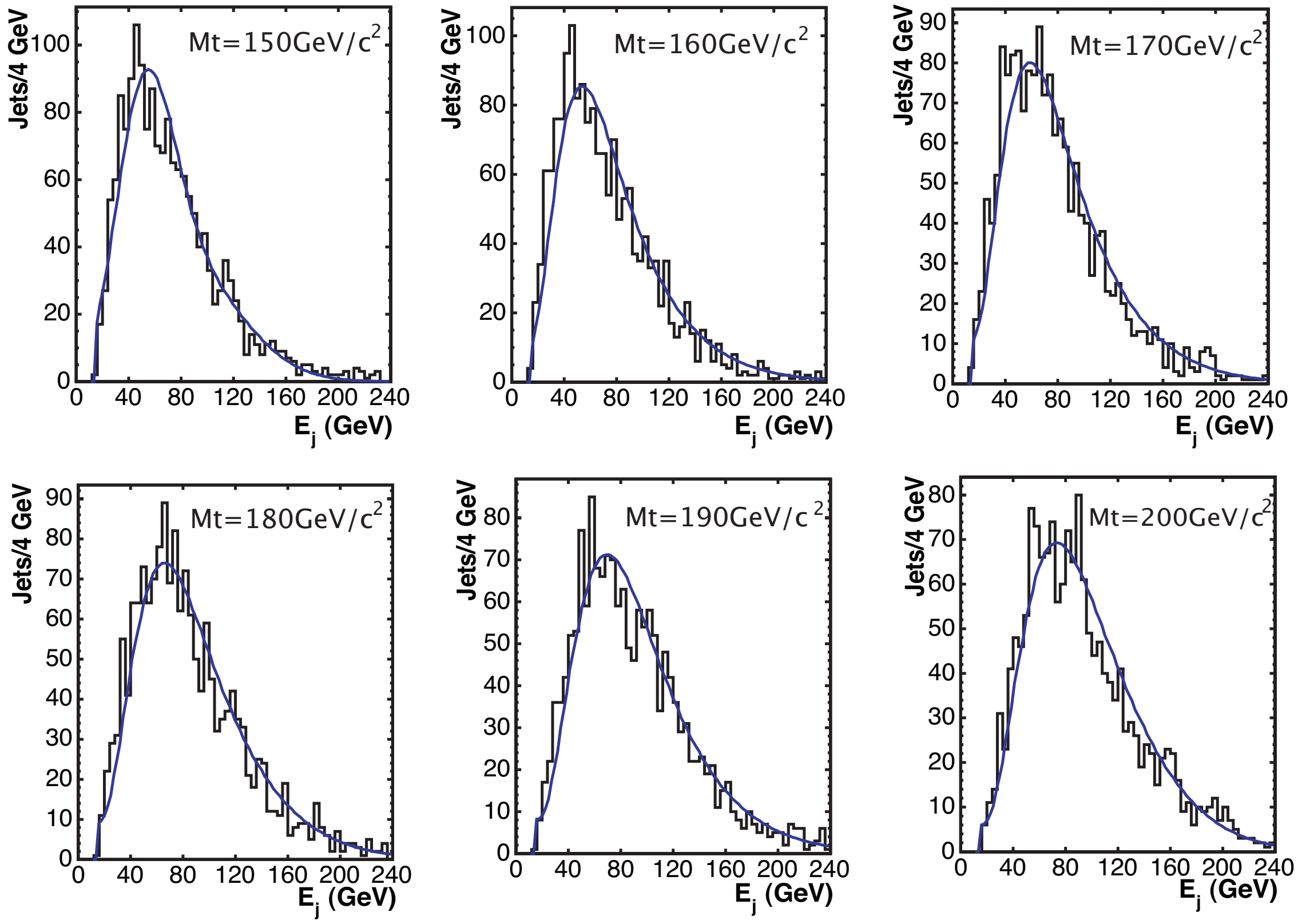

FIG. 3: Comparison of simulated $E_j$ with calculations from $W(E_p, E_j)$ from distributions $E_p$ in simulated samples with $M_t =$ 150, 160, 170, 180, 190, and 200 GeV/$c^2$. Histograms are simulated events for $M_t = 178$ GeV/$c^2$; lines show the calculated distributions using $W(E_p, E_j)$ derived using partons from $M_t = 178$ GeV/$c^2$.

While approximately 15% of $t\bar{t}$ pairs in $p\bar{p}$ collisions at $\sqrt{s} = 1.96$ TeV are produced in gluon-gluon fusion ($gg \to t\bar{t}$) [37], our studies have shown that this term can be excluded from the matrix element with very little loss of sensitivity to the measurement. A small systematic uncertainty is derived from theoretical uncertainty in the relative gluon fraction; see Section IX.C.

To evaluate the differential cross section with respect to observed quantities, $\frac{d\sigma}{d\mathbf{x}}$, we introduce conditional probability terms that relate the observed quantities to the parton level variables and subsequently integrate over unconstrained parton-level quantities, as described above.

The differential cross section is given by

$$\begin{aligned}\frac{d\sigma}{d\mathbf{x}} = &\left(\frac{1}{(2\pi)^3}\right)^6 \int \sum_{a,b} |M(q_1^i, q_2^j \to \ell_1, \nu_1, p_1, \ell_2, \nu_2, p_2)|^2 \\ &\times (2\pi)^4 \delta^4 (\sum_{k=1,2} (q_k - \ell_k - \nu_k - p_k)) \\ &\times \delta(q_{x_1})\, \delta(q_{y_1})\, dq_{x_1}\, dq_{y_1} \delta(q_{x_2})\, \delta(q_{y_2})\, dq_{x_2}\, dq_{y_2} \\ &\times \frac{(2\pi)^4 f^a_{\mathtt{PDF}}(q^i_{z_1}/E_{\mathtt{beam}})\; f^b_{\mathtt{PDF}}(q^j_{z_2}/E_{\mathtt{beam}}) dq_1^i\, dq_2^j}{4\sqrt{(q_1\cdot q_2)^2 - m^2_{q_1} m^2_{q_2}}} \\ &\times \delta^3(\ell_1 - L_1)\, \delta^3(\ell_2 - L_2)\, \frac{d^3\ell_1}{2E_{\ell_1}} \frac{d^3\ell_2}{2E_{\ell_2}} \\ &\times \delta(\theta_{p_1} - \theta_{j_1})\, \delta(\theta_{p_2} - \theta_{j_2}) \delta(\phi_{p_1} - \phi_{j_1})\, \delta(\phi_{p_2} - \phi_{j_2}) \\ &\times W(E_{p_1}, E_{j_1})\, W(E_{p_2}, E_{j_2}) \frac{d^3 p_1}{2E_{p_1}} \frac{d^3 p_2}{2E_{p_2}} \\ &\times f_{\mathtt{UTF}} \frac{d^3\nu_1}{2E_{\nu_1}} \frac{d^3\nu_2}{2E_{\nu_2}}.\end{aligned} \tag{13}$$

The sum is over possible incoming parton flavors. The term $1/4\sqrt{(q_1\cdot q_2)^2 - m^2_{q_1} m^2_{q_2}}$ reduces to $f_{\mathtt{flux}}(q_{z_1}, q_{z_2}) = 1/2 q_{z_1} q_{z_2}$ with the assumption that the parton mass is small in comparison to the longitudinal momentum.

The measured missing transverse energy is a sum of real missing transverse energy due to neutrinos, jet energy mismeasurement and unclustered energy from soft recoil which are not included in the reconstructed objects. The contributions from neutrinos and jet energy mismeasurement are accounted for by the matrix element and the transfer functions, respectively. Therefore, differences between the measured and solved (as below) missing transverse energy, $\not\!\!E_T$, are attributed to unclustered energy in the calorimeter. The $f_{\mathtt{UTF}}$ factor describes the prior probability of observing unclustered energy in the event; it is parametrized as a Gaussian in each of the $x$ and $y$ axes with no correlation. A width of 12 GeV/$c^2$ is extracted from a fit to simulated samples.

### A. Phase Space Transformation and Integration

We integrate over the lepton momenta, initial parton momenta, intermediate top quark and $W$ momenta, angular components of the $b$ partons and the six components of neutrino momenta.

In order to efficiently integrate over the parton-level variables, we perform a transformation which splits the original phase space into subspaces and introduces the equivalent number of extra variables and integrations. We introduce invariant masses that correspond to intermediate $t$ and $\bar{t}$ quarks and $W$ bosons. Each additional integration over an invariant mass of the intermediate particle has a corresponding $\delta$-function in squared invariant mass, and each intermediate particle four-momentum has corresponding $\delta^4$-function for the momentum conservation at the intermediate vertex.

The expression of the integrand is written in terms of the momenta of the final state particles, $b\ell\nu_\ell\bar{b}\ell'\bar{\nu}'_\ell$. Integration over the $t$ quark and $W$ boson invariant masses ($m_{t_i}$ and $m_{W_i}$) requires expressing the neutrino momenta in terms of these invariant masses. These two sets of variables are related by a system of six coupled quadratic equations written in terms of the final-state momenta and the $W$ boson momenta ($W_i$) and derived from expressions in the $\delta$ functions. Note that we explicity assume that the $t\bar{t}$ system has no transverse momentum and therefore do not use the measured missing tranverse energy to derive the solutions for neutrino energies. Solutions in which the measured missing transverse energy is significantly different from the sum of the neutrino tranverse energies are deweighted by the unclustered energy transfer function, $f_{\mathtt{UTF}}$.

$$\begin{aligned} m^2_{t_1} &= (p_1 + W_1)^2 \\ m^2_{t_2} &= (p_2 + W_2)^2 \\ m^2_{W_1} &= (\ell_1 + \nu_1)^2 \\ m^2_{W_2} &= (\ell_2 + \nu_2)^2 \\ (p_1 + \ell_1 + \nu_1 + p_2 + \ell_2 + \nu_2)_x &= 0 \\ (p_1 + \ell_1 + \nu_1 + p_2 + \ell_2 + \nu_2)_y &= 0 \end{aligned} \tag{14}$$

We rewrite these equations as a single fourth-order polynomial and find the solutions numerically using the Sturm Sequence approach [38]. The transformation between the phase space for neutrino momenta and invariant masses is not one-to-one due to the non-linearity of the relations. Multiple neutrino solutions may exist for specific invariant masses; such solutions are therefore summed. Other invariant masses may have no corresponding region of neutrino phase space and therefore no solutions and no contribution to the total probability.

Finally, the transformation of variables requires the inclusion of a Jacobian term, $J$. The final form of the

expression is

$$\frac{d\sigma}{d\mathbf{x}} = \left(\frac{1}{(2\pi)^3}\right)^6 (2\pi)^2 \int \sum_{a,b} \frac{|\mathcal{M}|^2}{2E^2_{\mathtt{beam}}} \frac{J^{-1}}{E_{\nu_1} E_{\nu_2} E_{L_1} E_{L_2}}$$
$$\times \frac{|p_1|^2 \sin\theta_{p_1} d|p_1|}{2E_{p_1}} \frac{|p_2|^2 \sin\theta_{p_2} d|p_2|}{2E_{p_2}}$$
$$\times f^a_{\mathtt{PDF}}\left(\frac{q_{z_1}}{E_{\mathtt{beam}}}\right) f^b_{\mathtt{PDF}}\left(\frac{q_{z_2}}{E_{\mathtt{beam}}}\right) f_{\mathtt{flux}}(q_{z_1}, q_{z_2})\ f_{\mathtt{UTF}}$$
$$\times W(E_{p_1}, E_{j_1})\ W(E_{p_2}, E_{j_2})$$
$$\times M_{t_1}\ M_{t_2}\ M_{W_1}\ M_{W_2}\ dM_{t_1}\ dM_{t_2}\ dM_{W_1}\ dM_{W_2}, \quad (15)$$

where the remaining integrations are over the invariant masses of the $t$ quarks and the $W$ bosons, and the magnitude of the $b$ quark momenta.

The cross section as a function of $M_t$ is expressed as a six-dimensional integral; this integration is performed numerically using the VEGAS[39] algorithm as implemented in the GNU Scientific Library[40].

## VII. BACKGROUND LIKELIHOODS

We calculate the per-event differential cross section for the three largest sources of background: $Z/\gamma^* + 2$ jet process ($Zjj$) where the $Z$ decays directly to electrons or muons, the $WW + 2$ jet process ($WWjj$) and the $W + 3$ jet processes ($Wjjj$) where one jet is misidentified as a lepton. $WZ$ with associated jets and $Z \to \tau\tau$ with two jets have a small overall contribution to the sample and are not directly modeled.

The major background processes can not be well described using a small number of diagrams. We therefore adapt routines from ALPGEN which make effective approximations to evaluate the matrix elements for these processes. The ALPGEN routines are a function of the spin and color configurations of the initial and final state partons as well as their momenta. We employ a statistical sampling over the spin and color configurations to numerically evaluate the averaged $\mathcal{M}$.

The final measurement is calibrated using fully realistic Monte Carlo events, which will incorporate the effects of these approximations.

### A. $Z/\gamma^* + 2$ jets

We employ the set of assumptions as described in Section IV and use transfer functions as defined and derived in Section V to connect the parton-level quantities to observed quantities. We use ALPGEN's $Z + 2p$ matrix element, which models the associated production of two light quark $(u, d, c, s)$ and gluon jets. As the number of unmeasured quantities is fewer than in the case of the signal likelihood, we can relax the assumption that the $p_T$ of the $Zjj$ system is zero and instead integrate over the unknown $p_T$ which arises from additional softer jets and unclustered energy. We express this as $r_x$ and $r_y$, components of the recoil in the $x$ and $y$ axes, respectively. We integrate over $dr_x dr_y$ using uncorrelated Gaussian priors in $x$ and $y$ with widths of 12 GeV/$c^2$, $f_{UTF}(r_x, r_y)$, as extracted from simulated samples. The differential cross section can be expressed as

$$\frac{d\sigma_{Zjj}}{d\mathbf{x}} = \frac{1}{(2\pi)^8} \frac{1}{16^2} \int \sum_{i,j} dr_x dr_y f_{UTF}(r_x, r_y)$$
$$\times \frac{f^i_{\mathtt{PDF}}(q_{z_1}/E_{\mathtt{beam}})\ f^j_{\mathtt{PDF}}(q_{z_2}/E_{\mathtt{beam}})}{E_{L_1} E_{L_2} E_{p_1} E_{p_2}} \frac{|\mathcal{M}|^2}{2} \quad (16)$$
$$\times \frac{W(E_{p_1}, E_{j_1})\ W(E_{p_2}, E_{j_2})\ |p_1|^2\ |p_2|^2}{|q_{z_1} q_{z_2}|\ |\sin(\phi_{j_1} - \phi_{j_2})|},$$

where $p_1, p_2$ are the four-momenta of the final state partons which lead to creation of extra jets, $L_1, L_2$ are the four-momenta of the final state leptons, and $q_1, q_2$ are the four-momenta of incoming partons.

### B. $WW + 2$ jets

The production of $W$ pairs with associated jets is modeled in a similar fashion. We use ALPGEN's $WW + 2p$ matrix element, which models the associated production of two light quark $(u, d, c, s)$ and gluon jets. After making the assumptions listed in Section IV, we choose to transform the phase-space by introducing the invariant masses of the intermediate $W$ bosons. We express all the parton variables except the neutrino momenta in spherical coordinates. Integrating over delta functions for lepton energy, jet angles, and total conservation of momentum gives

$$\frac{d\sigma_{WWjj}}{d\mathbf{x}} = \frac{1}{(2\pi)^{16}} \int \sum_{a,b} \frac{|\mathcal{M}|^2}{2E^2_{\mathtt{beam}}}$$
$$\times \frac{|p_1|^2 \sin\theta_{p_1} d|p_1|}{2E_{p_1}} \frac{|p_2|^2 \sin\theta_{p_2} d|p_2|}{2E_{p_2}}$$
$$\times \frac{J^{-1}}{E_{\nu_1} E_{\nu_2} E_{L_1} E_{L_2}} f^a_{\mathtt{PDF}}\left(\frac{q_{z_1}}{E_{\mathtt{beam}}}\right) f^b_{\mathtt{PDF}}\left(\frac{q_{z_2}}{E_{\mathtt{beam}}}\right)$$
$$\times f_{\mathtt{flux}}(q_{z_1}, q_{z_2}) W(E_{p_1}, E_{j_1})\ W(E_{p_2}, E_{j_2})$$
$$\times M_{W_1}\ M_{W_2}\ dM_{W_1}\ dM_{W_2} d\nu_{1z} d\nu_{2z}, \quad (17)$$

where $L_1$, $L_2$ are the measured four-momenta , $\ell_1$, $\ell_2$ are the parton-level four-momenta of the final state leptons, $\nu_1$, $\nu_2$ are the four-momenta of the final state neutrinos, $p_1$, $p_2$ are the four-momenta of the final state partons that lead to creating extra jets, and $q_1$, $q_2$ are the four-momenta of the incoming partons. The sum runs over the incoming parton flavors.

The final integration is performed over the the momenta of the partons which lead to jet production, the $W$ boson invariant masses, and the $z$ components of neutrino momenta. Transformation of the space requires solving a coupled system of equations to express the neutrino energies in terms of the $W$ masses.

### C. Modeling of Backgrounds with Jets Misidentified as Leptons

Events in which a jet is misidentified as a lepton can be modelled with the process $p\overline{p} \to Wjjj \to l\nu jjj$. We use ALPGEN's $W+3p$ matrix element, which models the production of three light quark $(u,d,c,s)$ and gluon jets. Using this process as the basis for the model, we sum over the probability that either reconstructed lepton is a misidentified jet. Calculation of the cross-section follows the style of the other calculations above.

Making the standard assumptions, and integrating over all delta functions gives

$$\begin{aligned}\frac{d\sigma_{Wjjj}}{d\mathbf{x}} = &\int \sum_{a,b} \frac{d|p_1|d|p_2|d|p_3|d\nu_z}{(2\pi)^8 32 E_L E_\nu E_{p_1} E_{p_2} E_{p_3}} \\ &\times W(E_{p_1},E_{j_1})W(E_{p_2},E_{j_2})W(E_{p_3},E_{j_3}) \\ &\times |p_1|^2|p_2|^2|p_3|^2 \sin\theta_{j_1}\sin\theta_{j_2}\sin\theta_{j_3} \\ &\times f^a_{\mathtt{PDF}}\left(\frac{q_{z_1}}{E_{\mathtt{beam}}}\right) f^b_{\mathtt{PDF}}\left(\frac{q_{z_2}}{E_{\mathtt{beam}}}\right) f_{\mathtt{flux}}(q_{z_1},q_{z_2})\frac{|\mathcal{M}|^2}{2},\end{aligned} \tag{18}$$

where $p_1, p_2, p_3$ are the four-momenta of the final state partons which lead to creation of extra jets, $L$ is the four-momenta of the final state lepton, $\nu$ the four-momentum of the final state neutrino, and $q_1, q_2$ are the four-momenta of incoming partons. Jets which are misidentified as a lepton have the large majority of their momentum carried by a single leading object ($\pi^0$ or $\pi^\pm$); therefore, we assume that the jet identified as a lepton carries the momentum of the parton.

## VIII. TOP QUARK MASS RECONSTRUCTION

In this section, we discuss the combination of the per-event differential cross section calculations for signal and background into a joint probability for a sample of events, the procedure for mass extraction, and calibration of the method using simulated events.

### A. Posterior Probability

We express the individual event probability density in $\mathbf{x}$ as a sum of the signal and background probabilities with their respective fractions, see Eq. 4. The signal and background fractions are expressed in terms of the number of expected signal events $\lambda_s(M_t)$ and background $\lambda_b$ events:

$$\begin{aligned} p_s(M_t) &= \frac{\lambda_s(M_t)}{\lambda_s(M_t)+\lambda_b} \\ p_{b_i}(M_t) &= \frac{\lambda_{b_i}}{\lambda_s(M_t)+\lambda_b}, \end{aligned} \tag{19}$$

where $\lambda_b = \sum_i \lambda_{b_i}$. The expected signal $\lambda_s(M_t)$ is calculated relative to a reference point $M_0$ and extrapolated to other masses using the mass dependence of the total accepted cross section $\sigma_s(M_t)\epsilon_s(M_t)$:

$$\lambda_s(M_t) = \lambda_s(m_0)\frac{\sigma_s(M_t)\epsilon_s(M_t)}{\sigma_s(M_0)\epsilon_s(M_0)}, \tag{20}$$

where $\sigma_s(M_t)$ is the total production cross section [31] and $\epsilon_s(M_t)$ is the acceptance measured in Monte Carlo events.

For an individual event $\mathbf{x}$, $P(\mathbf{x}|M_t)$ is a likelihood in $M_t$. The posterior probability density in $M_t$ is the product of a flat prior probability and the product of the individual event likelihoods. The measured mass, $\hat{M}_t$, is chosen as the expectation value of the posterior probability to avoid potential fluctuations in the position of the maximum probability:

$$\hat{M}_t = \frac{\int dM_t M_t P(\mathbf{x}|M_t)}{\int dM_t P(\mathbf{x}|M_t)} \tag{21}$$

The measured statistical uncertainty, $\Delta\hat{M}_t$, is the standard deviation of the posterior probability,

$$\Delta\hat{M}_t = \sqrt{\frac{\int dM_t M_t^2 P(\mathbf{x}|M_t)}{\int dM_t P(\mathbf{x}|M_t)} - (\hat{M}_t)^2} \tag{22}$$

### B. Calibration

We have searched for any potential biases on the extracted mass or its uncertainty due to our fitting procedure. We parameterize this bias using a linear correction factor to the measured mass consisting of an offset $M_0$ and a slope $s_{M_t}$,

$$M_t = 178.0\,\mathrm{GeV}/c^2 + (\hat{M}_t - M_0)/s_{M_t}, \tag{23}$$

and a simple scale factor,

$$\Delta M_t = S_\Delta \times \Delta\hat{M}_t/s_{M_t}, \tag{24}$$

to the measured statistical error in data, where $M_0, s_{M_t}, S_\Delta$ are extracted from ensembles of Monte Carlo experiments. In the following sections, we study the calibration of the method in Monte Carlo experiments with only $t\bar{t}$ events as well as in Monte Carlo experiments with both $t\bar{t}$ and background events.

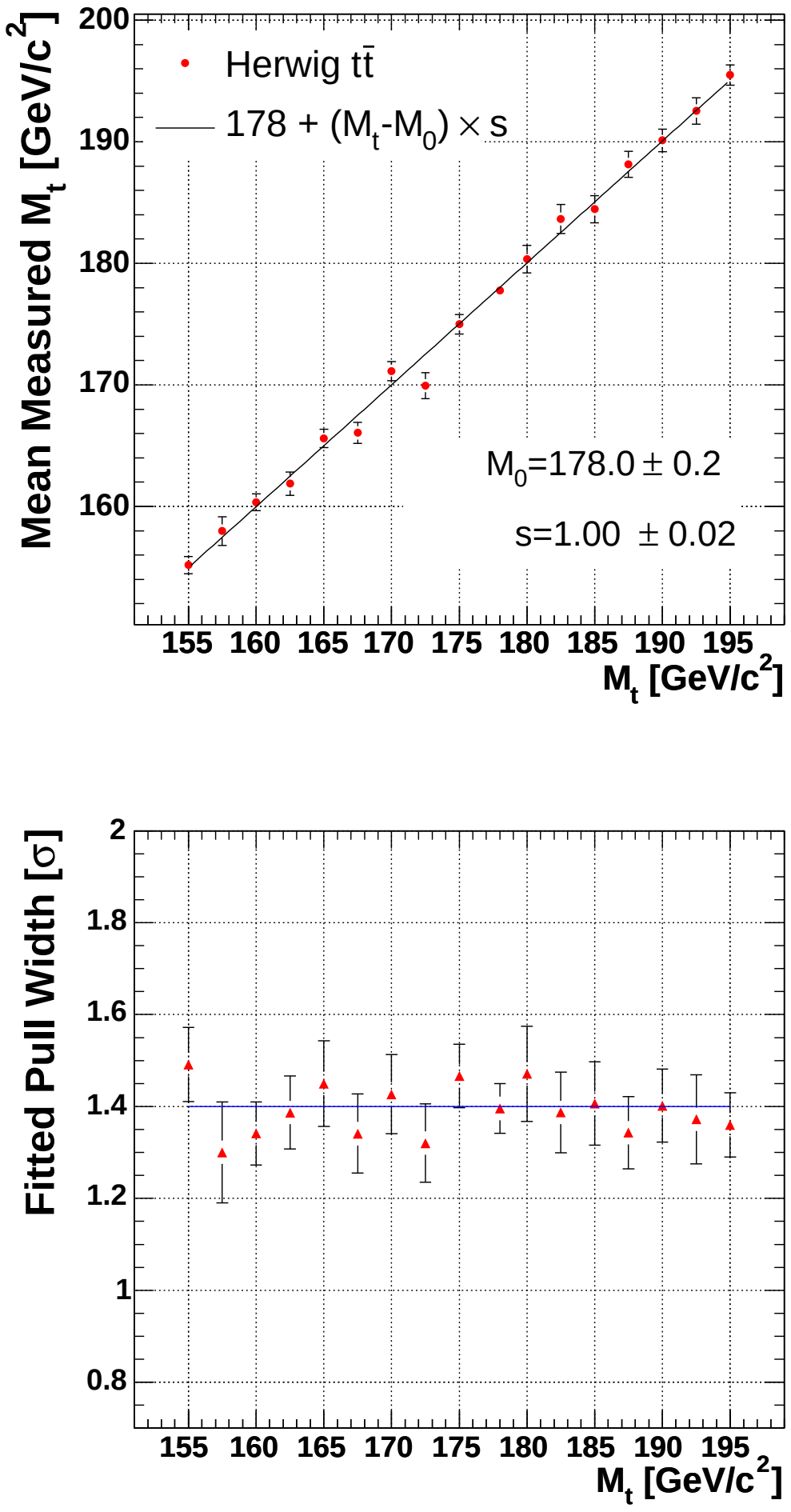

FIG. 4: Top, mean of measured $M_t$ in Monte Carlo experiments with only $t\bar{t}$ events. Bottom, fitted widths of pull distributions from the same Monte Carlo experiments.

## C. Monte Carlo Experiments with $t\bar{t}$ Events

We first construct Monte Carlo experiments of $t\bar{t}$ events with varying $M_t$, each generated by HERWIG [34], and a simulation of the CDF II detector. The Monte Carlo experiments contain a number of $t\bar{t}$ events drawn from a Poisson distribution whose mean corresponds to the number of $t\bar{t}$ events expected in a sample at the given mass. No background events are included in these experiments, and the $p_{b_i}$ in Eq. 4 are therefore identically zero. Figure 4 shows the mean measured $M_t$ and the width of the pull distribution at each mass, where a pull is defined as $(M_t - \hat{M}_t)/\Delta M_t$ for each Monte Carlo experiment.

The mean measured top quark mass in these Monte Carlo experiments shows no evidence of bias. A linear fit to the extracted mass yields $M_0 = 178.0$ GeV/$c^2$ and $s_{M_t} = 1.00$. The pull width, however, indicates that the extracted statistical error is consistently underestimated by a factor of $S_\Delta = 1.4$, independent of top quark mass. This is due to the several simplifying assumptions made in the formulation of the probability expression. As a result, it does not contain a description of some effects which cause smearing of the extracted mass. The most important of these are the assumption that all jets come from $b$ quarks, rather than from initial or final state QCD radiation, that jet angles accurately give the parton angles, and that lepton energies are perfectly measured. We use Monte Carlo experiments to measure the effects of these simplifying assumptions on the measured top quark mass. This is done by measuring our ability to extract the top quark mass in Monte Carlo experiments where these assumptions are violated to increasing degrees.

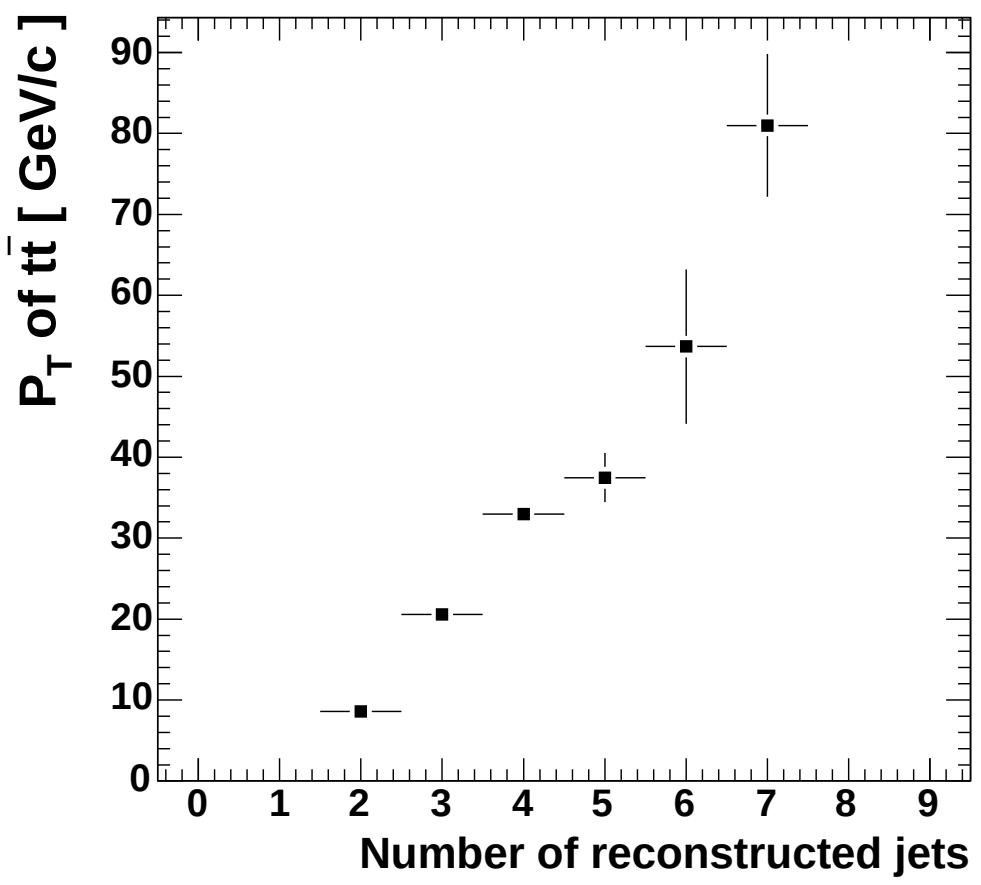

FIG. 5: Correlation between the mean $p_T$ of the $t\bar{t}$ system and the number of reconstructed jets in PYTHIA simulated $t\bar{t}$ events.

### 1. *Jet-Parton Assignment*

In samples of simulated events which pass selection requirements, 70% of events contain two reconstructed jets whose axes lie within a cone of $\Delta R \equiv \sqrt{(\Delta\eta)^2 + (\Delta\phi)^2} < 0.7$ of unique $b$ quarks from the top quark decay. Monte Carlo experiments using this subset of events have a significantly smaller pull width, 1.20. This suggests that events in which the assumption of correspondence between jets and $b$ quarks is violated contribute significantly to non-unit pull widths.

We note that the largest source of incorrectly assigned jets is initial state radiation, where a hard emission from the incoming quark or gluon gives rise to a jet. In simulated events, a strong correlation is seen between the number of reconstructed jets and the $p_T$ of the $t\bar{t}$ system, which is a measure of the total recoil against initial state radiation; see Figure 5.

### 2. Lepton Resolution

Though lepton energies are well measured by CDF, electrons and muons are measured by different subdetectors. The energy of electrons at high $E_T$ is very well measured by the calorimeter [14]:

$$\frac{\sigma_E}{E} = \frac{13.5\%}{\sqrt{E(\mathrm{GeV})}} \oplus 2.0\%. \quad (25)$$

The momentum of muons is measured by the central tracker, whose resolution begins to degrade at large $p_T$ [14]:

$$\frac{\sigma_{p_T}}{p_T} = 0.0011 \cdot p_T(\mathrm{GeV}/c). \quad (26)$$

Monte Carlo experiments formed using events in which jets are matched well to $b$ quarks and events containing only electrons have a pull width of 1.10. Similar Monte Carlo experiments using only muons have a pull width of 1.20, indicating that muon momentum resolution contributes to pull widths greater than unity. Electrons and muons which arise from $W \to \tau\nu_\tau$ decays are not well described by the matrix-element so they contribute to the pull width as well.

### 3. Jet Angle Resolution

The jet angle resolution is finite, though it is significantly more precise in comparison to the jet energy resolution. Figure 6 shows the angular distance, $R \equiv \sqrt{(\Delta\eta)^2 + (\Delta\phi)^2}$, between reconstructed jets and the closest $b$ quarks in fully simulated $t\bar{t}$ events passing the selection criteria in Section III. Jets which are not matched to either of the $b$ quarks from top quark decay (no $b$ quark within $R < 0.7$) are likely due to initial state radiation, as described above.

To isolate the effect of the jet-angle resolution, we examine a subset of events from a fully simulated sample with $M_t = 178$ GeV/$c^2$. To remove the effects of lepton resolution and jet-parton matching as isolated above, we require well measured leptons ($p_T^{lepton} - p_T^{reconstructed} < 2$ GeV/$c$) and matched jets ($R < 0.4$). Events of this type have a negligible rate of jet-parton misassignment. Monte Carlo experiments with this subset of events have a pull width of 1.1, consistent with experiments described above which use only electrons and matched jets. Further tightening the $R$ requirement reduces the pull width to 1.0.

The effect on the pull width of jet angle resolution is confirmed by observation of a $\approx$ 10% increase in pull width when jet angles are smeared in otherwise perfectly measured parton-level events.

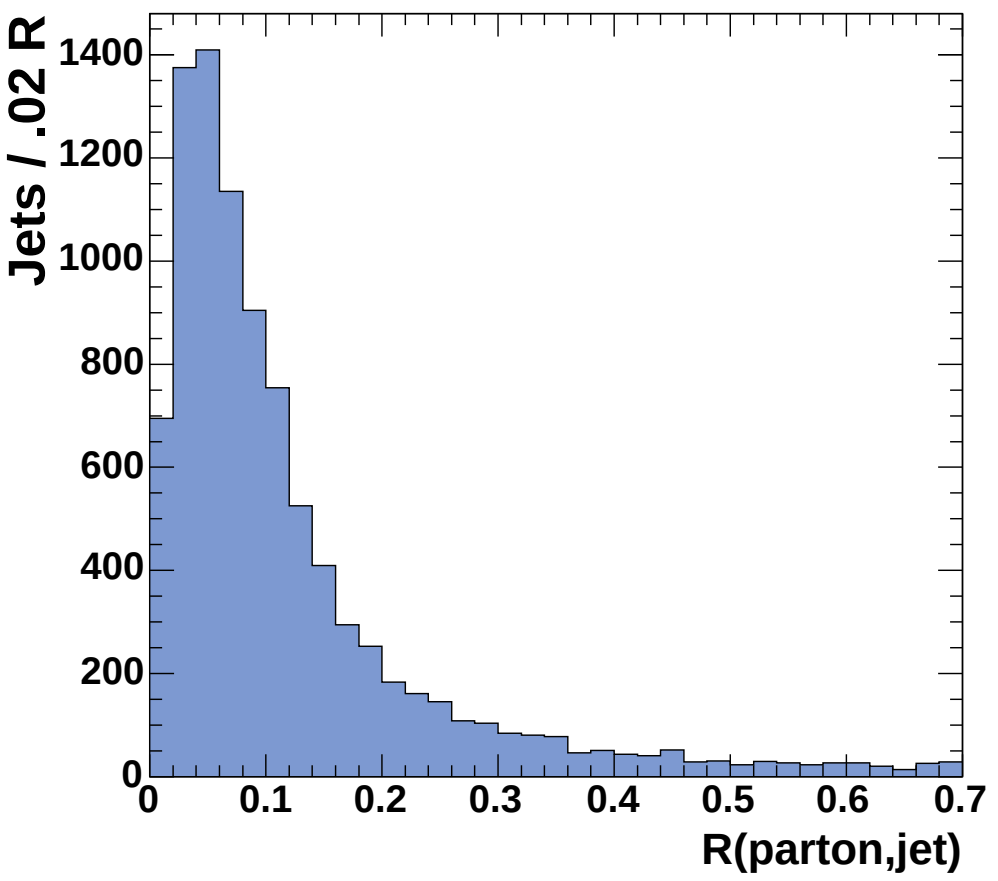

FIG. 6: Angular distance, $R$, between a reconstructed jet and the closest $b$ quark in $t\bar{t}$ events simulated with HERWIG. The width of the distribution demonstrates the angular resolution. 20% of jet-parton pairs have $R \geq 0.7$ (not shown), coming from jets with no corresponding $b$ quark.

## D. Monte Carlo experiments with $t\bar{t}$ events and background events

To more fully describe the sample of events we expect in the data, we construct Monte Carlo experiments with $t\bar{t}$ events as well as events which model the background sources listed in Table I. Each Monte Carlo experiment contains numbers of events drawn from Poisson distributions whose means are given by the expected number of events from each given source.

To extract the mass, we use the full probability expression of Eq. 4, including terms which model the production of three background sources. Figure 7 shows the mean extracted mass and width of pull distributions in these Monte Carlo experiments. A linear fit to the extracted mass yields

$$M_0 = 177.2 \pm 0.21 \text{ GeV}/c^2, s_{M_t} = 0.84 \pm 0.02$$

as defined in Eq. 23. These parameters indicate a small bias due to lack of modeling of the minor backgrounds ($WZ$ and $Z \to \tau\tau$) and imperfect descriptions of major backgrounds. If the background probability calculations are not included, $M_0$ becomes 175.8 and $s_{M_t}$ decreases to 0.72. This demonstrates that the inclusion of the background probabilities does improve the sensitivity of the method. The width of the pull distributions in fully realistic Monte Carlo experiments using the full probability expression is fit to a straight line to extract the error scale factor,

$$S_\Delta = 1.51 \pm 0.02.$$

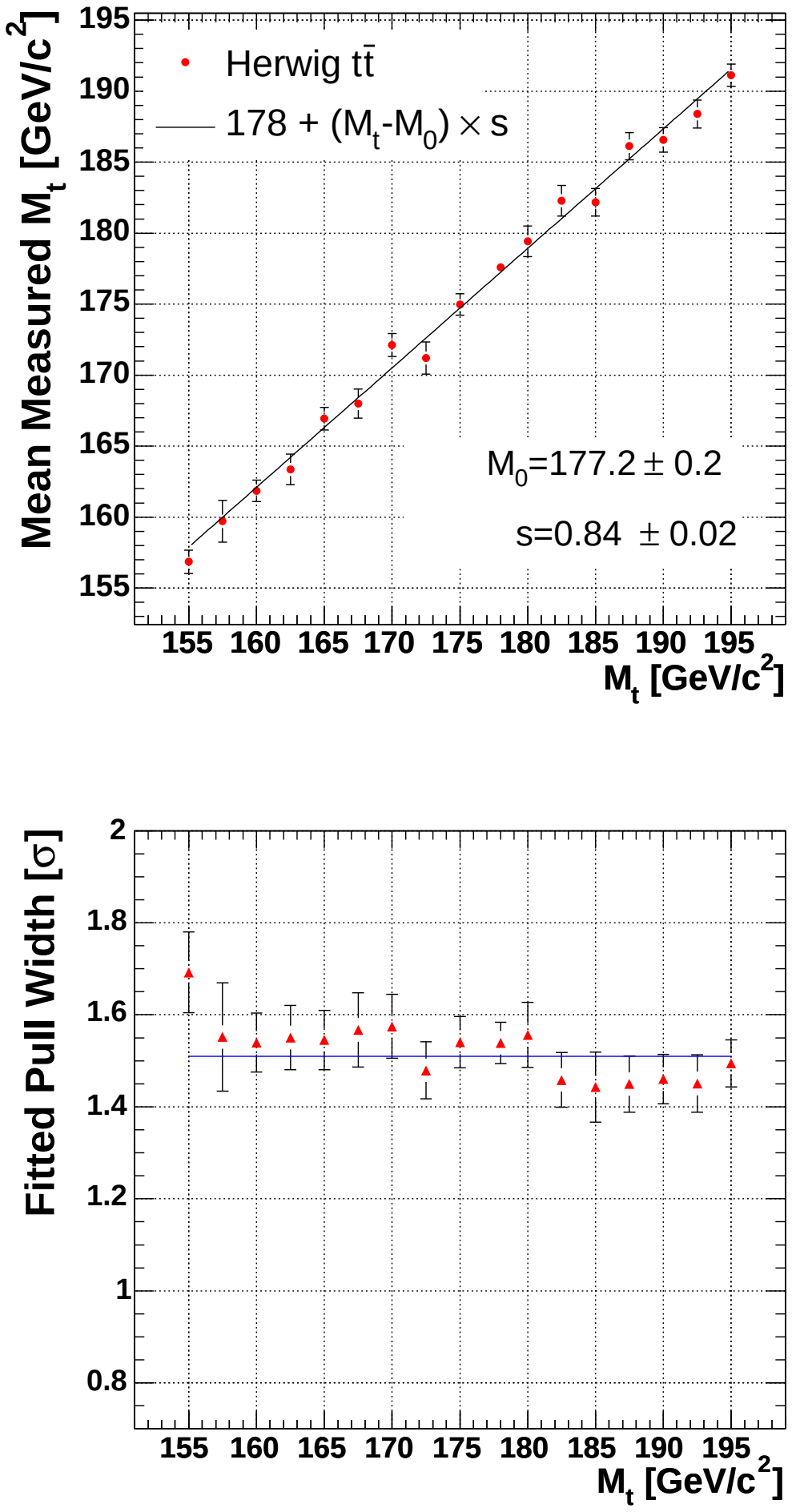

FIG. 7: Top, mean of measured $M_t$ in Monte Carlo experiments with $t\bar{t}$ events as well as background events. Bottom, fitted widths of pull distributions from the same Monte Carlo experiments.

This scale factor is larger than the scale factor of $S_\Delta = 1.4$ measured in experiments with no background events. This is due to the presence of unmodeled background events ($WZ$ and $Z \to \tau\tau$), which smear the extracted mass without inflating the measured statistical error.

Uncertainty on the measured scale factor would contribute a systematic error to the statistical error of the measurement. The measured scale factor is stable to within 5% under variations of the sources of systematic uncertainty in the modeling of the calibration Monte Carlo described in Section IX below.

After application of mass and error corrections, the final measured mass is unbiased and the pull distribution is well described by a Gaussian; see Figures 8 and 9.

## IX. SYSTEMATIC UNCERTAINTIES

The measurement is calibrated using Monte Carlo simulated events. Therefore, the majority of systematic uncertainties come from uncertainties in this modeling of the data. In this Section, we describe the facets of the simulation which may not accurately describe the observed data and estimate the effect on the measurement.

To measure the size of the impact of each uncertainty, we perform Monte Carlo experiments using a pool of simulated events in which a feature of the events has been modified. By extracting the average measured mass, we can determine the typical shift due to these features.

### A. Systematic Uncertainties Due to Jet Energy Scale

The largest source of systematic uncertainty arises from potential mismodeling of the jet energy measurement, through uncertainties in the various corrections applied to the measured jet energy [28]. These jet energy corrections involve knowledge of the absolute energy scale, energy loss outside the jet search cone $\Delta R$, the non-uniformity in response of the calorimeter as a function of $\eta$, effects from multiple $p\bar{p}$ collisions and energy deposition from the underlying $p\bar{p}$ event. These uncertainties are shown as a function of jet $P_T$ in Figure 10. A systematic uncertainty is estimated for each jet energy correction by performing Monte Carlo experiments drawn from simulated signal and background events with $\pm 1$ standard deviation in correction uncertainty, and taking the half-difference in mean reconstructed top quark mass between the two results. The uncertainties from each energy correction are then added in quadrature to arrive at a total systematic uncertainty on the jet energy scale. No strong dependence was observed as a function of the top quark mass.

Since the above jet energy corrections are developed from studies of samples dominated by light quark and gluon jets, additional uncertainty occurs from extrapolating this procedure to $b$ quarks. The resulting systematic effect on jet energy is considered to stem from three main sources: uncertainty in the $b$ jet fragmentation model, differences in the energy response due to semi-leptonic decays of $b$ hadrons, and uncertainty in the color flow within top quark production and decay to $b$ jets [41]. As in the jet energy scale uncertainty, Monte Carlo experiments are performed on events where the $b$ jet energies have been altered by $\pm 1$ standard deviation for each uncertainty, and the resulting half-differences added in quadrature to estimate the total systematic uncertainty due to $b$ jet energy uncertainty.

Table III lists the uncertainties due to each correction. The sum in quadrature of the uncertainties in Table III yields a final jet energy scale systematic uncertainty of 2.6 GeV/$c^2$.

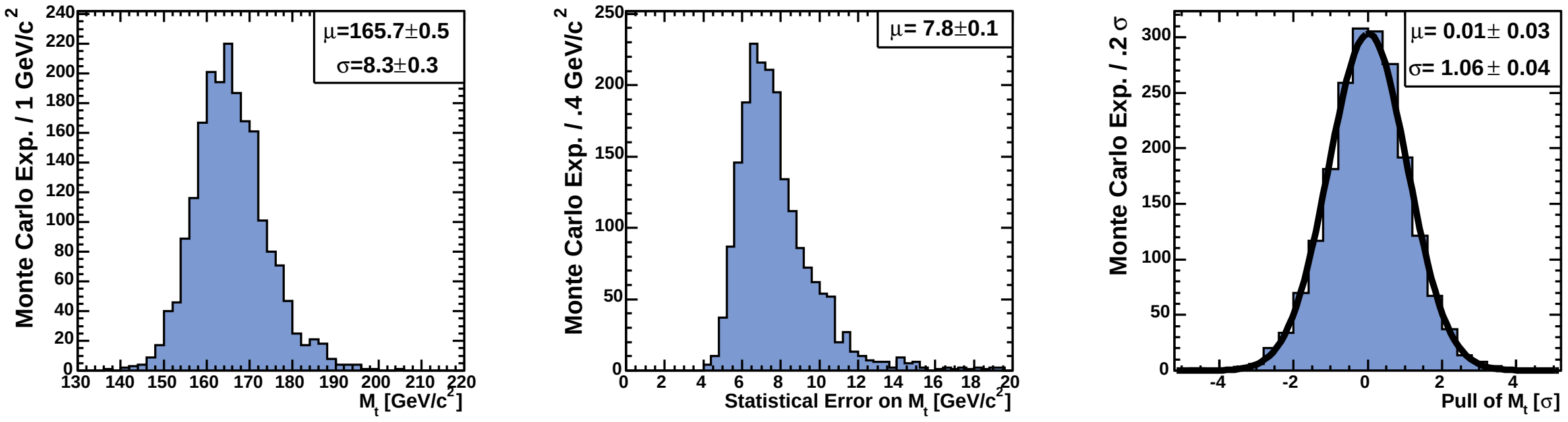

FIG. 8: Left, distribution of measured $M_t$ in Monte Carlo experiments with $t\bar{t}$ events at $M_t = 165$ GeV/$c^2$ and background events. Measured statistical error, center, and pull distribution, right, are also shown. All mass and error corrections have been applied.

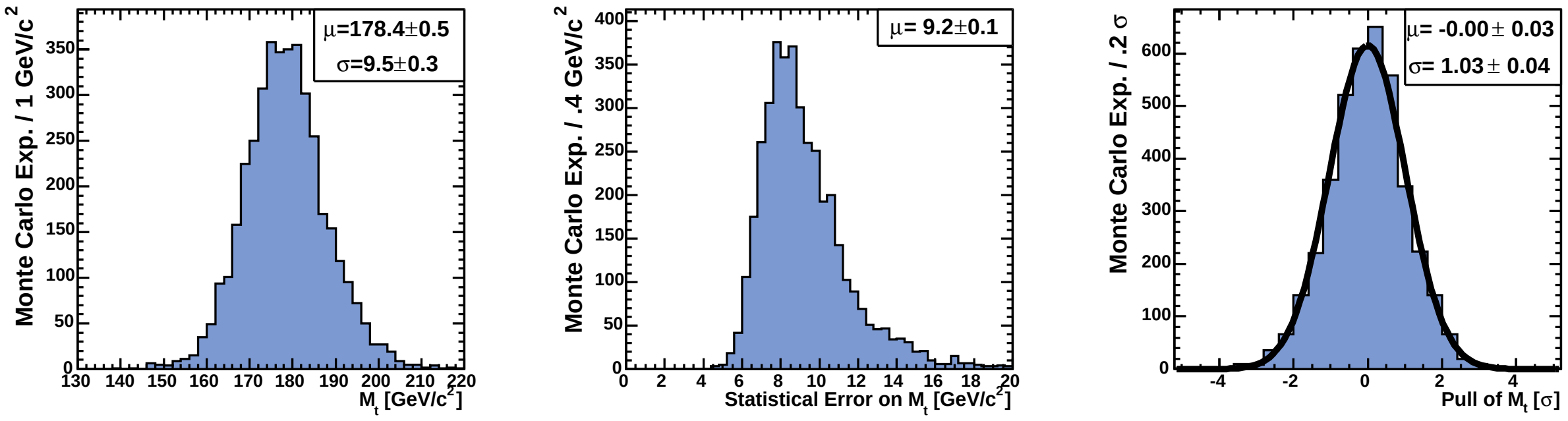

FIG. 9: Left, distribution of measured $M_t$ in Monte Carlo experiments with $t\bar{t}$ events at $M_t = 178$ GeV/$c^2$ and background events. Measured statistical error, center, and pull distribution, right, are also shown. All mass and error corrections have been applied.

TABLE III: Systematic uncertainties on the top quark mass due to each jet energy systematic uncertainty.

| Jet Energy Systematic | $\delta M_t$ (GeV/$c^2$) |
|---|---|
| Modeling of hadron jets (absolute scale) | 1.1 |
| Modeling of parton showers (out-of-cone) | 2.2 |
| Response relative to central calorimeter | 0.8 |
| Underlying event and multiple interactions | 0.1 |
| Modeling of $b$ jets | 0.5 |
| Total jet energy systematic uncertainty | 2.6 |

## B. Systematic Uncertainties Due to Backgrounds

There are two major sources of systematic uncertainty that result from the presence of background. One is the limited number of events to model the background and calibrate Monte Carlo experiments. The other is the uncertainty in modeling major sources of background.

### 1. *Background Statistics*

We estimate the sensivity of the measurement to the limited number of events available to model the background processes in the calibrating Monte Carlo experiments. We perform Monte Carlo experiments using disjoint subsets of statistically independent event samples for each background process. The root mean square of the difference between the mass measured in each pair of experiments is an estimate of this uncertainty and is measured to be 1.2 GeV/$c^2$, driven largely by the limited amount of data available for modeling events with a jet misidentified as a lepton.

### 2. *Background Modeling*

We estimate the sensitivity to the modeling of the two largest backgrounds, Drell-Yan and misidentified leptons.

The contribution to the sample from Drell-Yan production comes from events in which there is an apparently large missing transverse energy due to mismeasurement. To gauge the sensitivity to events on the tail of the distribution, we vary the composition of the Monte Carlo experiments by either enhancing events or suppressing events on the tail. We assign weights to the events proportional to the measured missing transverse energy, and to its inverse:

$$w_+ \propto \not{E}_T, \qquad w_0 = 1, \qquad w_- \propto \not{E}_T^{\,-1}.$$

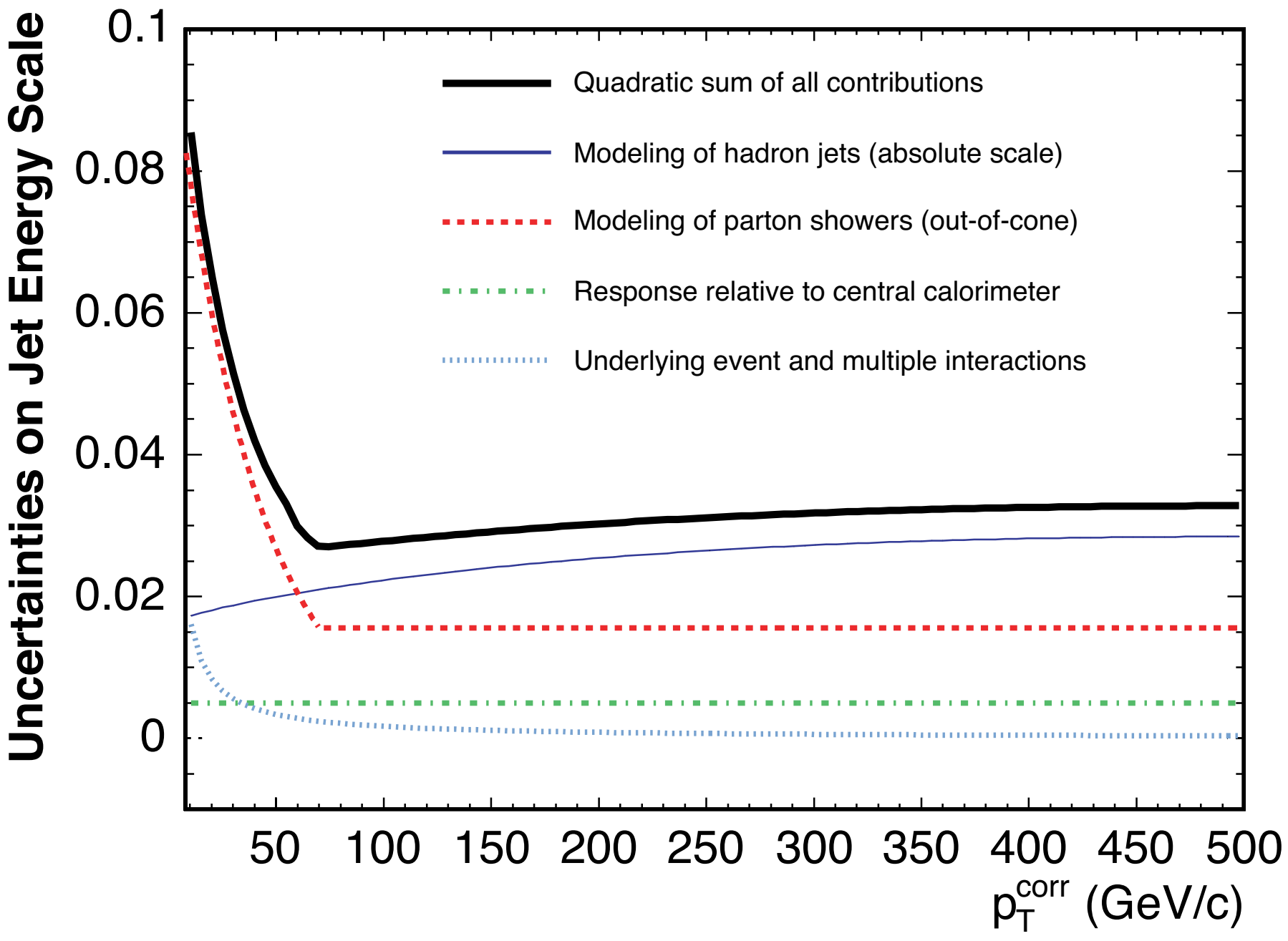

FIG. 10: Systematic uncertainties for several sources in the jet energy scale, as a function of the corrected $p_T$ of jets ($p_T^{corr}$) in the central ($0.2 < |\eta| < 0.6$) calorimeter.

This prescription conservatively describes the uncertainty in modeling of the missing transverse energy in events from Drell-Yan production, see Figure 11. To determine the systematic uncertainty arising from the shape of the Drell-Yan background, we compare the mass resulting from Monte Carlo experiments using $w_+$ and $w_-$ weights to that obtained using unit weight ($w_0$). The larger of the two differences is taken to be the systematic uncertainty and is measured to be 0.4 GeV/$c^2$.

The background resulting from events in which a jet is misidentified as a lepton is very difficult to describe accurately with simulation, as it is very sensitive to the details of the detector performance. To avoid issues of modeling, the events which imitate this background are drawn from the data themselves. These events are selected with the same requirements as for the candidates, except for a looser requirement on one of the leptons, and then are weighted by the probability that the loose lepton would pass lepton identification requirements. These weights are measured as a function of the $p_T$ and isolation of the fake candidate using a sample dominated by jets; each candidate has its own weight ($w$) and uncertainty ($\Delta w$). To gauge the sensitivity to the calculation of the fake rates, we vary the fake rates in two ways. First ($a$), we enhance those events with fake probability greater than the mean ($w > \overline{w}$) to exaggerate their effect; second ($b$), we enhance events with fake probability smaller than the mean($w < \overline{w}$), to exaggerate their effect:

| Mode | $w > \overline{w}$ | $w < \overline{w}$ |
|---|---|---|
| ($a$) | $w \to w + \Delta w$ | $w \to w - \Delta w$ |
| ($b$) | $w \to w - \Delta w$ | $w \to w + \Delta w$ |

We take the difference between the mass obtained using ($a$) and that obtained using ($b$) as the systematic uncertainty, 0.6 GeV/$c^2$, and the total shape uncertainty to be the two (Drell-Yan and fakes) summed in quadrature, 0.8 GeV/$c^2$.

### C. Parton Distribution Function Uncertainties

To propagate uncertainties in the parton distribution functions (PDFs) to the mass measurement, we reweight a sample of simulated PYTHIA events according to 20 sets of $\pm 1\sigma$ uncertainty eigenvectors based on CTEQ6M [42]. The uncertainty using each of the eigenvectors is added according to the standard prescription to yield the total uncertainty of $\Delta_{PDF} = 0.55$ GeV/$c^2$.

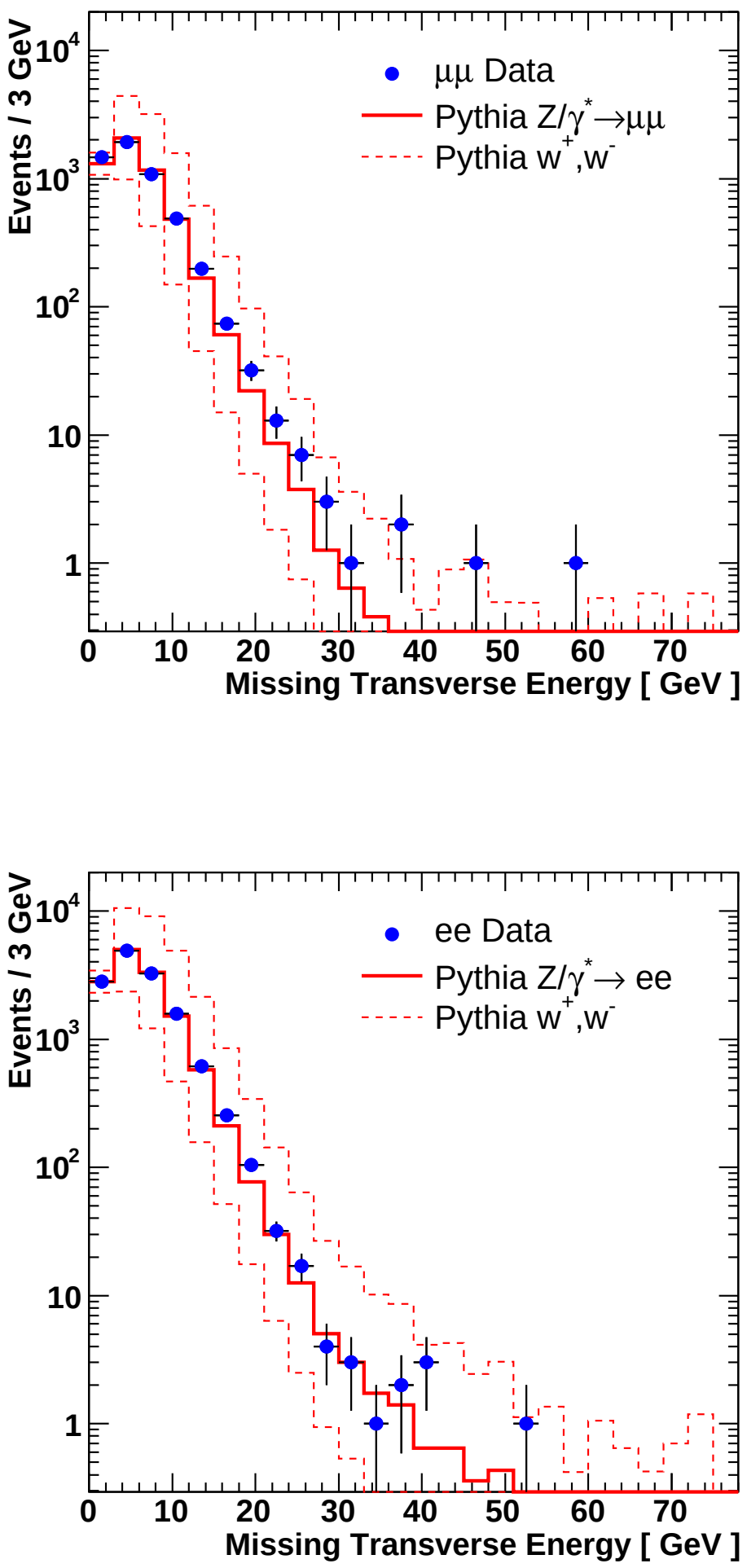

FIG. 11: Comparison of the missing transverse energy in dilepton events in data, PYTHIA simulated $Z/\gamma^* \to ll$ events and simulated events weighted as described in the text to enhance or suppress events with large missing transverse energy. Top shows $\mu\mu$ data and bottom shows $ee$ data.

The two samples MRST72 and MRST75 use different values of $\Lambda_{QCD}$ (228 and 300 MeV) to calculate the PDFs. The difference in the extracted mass from Monte Carlo experiments using these two samples provides a measure of the sensitivity to the uncertainty in $\Lambda_{QCD}$. This uncertainty is measured to be 0.85 GeV/$c^2$.

The leading order PYTHIA and HERWIG generators contain $\sim$5% $gg$ events in their initial state. In $p\bar{p}$ collisions at $\sqrt{s} = 1.96$ TeV , approximately 15% of the $t\bar{t}$ pairs are produced from $gg$ annihilation. The matrix element used in the likelihood calculation describes $t\bar{t}$ production from $q\overline{q}$ annihilation only. To measure the sensitivity to the initial state, we vary the fraction of $gg$ initial states in the Monte Carlo experiments from 0% to 15%. We take the difference of the measured mass using 0% and 15% $gg$ events as a systematic uncertainty, which is measured to be 0.35 GeV/$c^2$.

The total PDF uncertainty is obtained by adding the above three uncertainties in quadrature, and is 1.1 GeV/$c^2$.

### D. Systematic Uncertainties Due to Initial and Final State Radiation

The calculation of the signal probability does not contain a description of initial (ISR) or final state radiation (FSR), which may contribute significantly to the fraction of mismeasured events. The rate of additional jets from initial state radiation can be estimated in $Z$+jets events, as there are no significant contributions from final state radiation; the radiation is found to depend smoothly on the Drell-Yan mass squared [41], and can be examined over a broad range of energies, extending up to the range of $t\bar{t}$ production. To measure the uncertainty due to imperfect knowledge of the rate of radiation, we examine the measured top quark mass in samples where the Monte Carlo generator parameters are varied by very conservative amounts based on the studies in the $Z$+jets events. We measure the uncertainty due to ISR to be 0.5 GeV/$c^2$.

Final state radiation can be probed in the same manner, as it is described by the same showering algorithm. We find the uncertainty due to FSR to be 0.5 GeV/$c^2$.

### E. Other Systematic Uncertainties

To account for a possible bias of the Monte Carlo modeling, we measure the difference in top quark mass as extracted from HERWIG and PYTHIA samples. This amounts to 0.8 GeV/$c^2$.

In addition, uncertainties in the fitted response calibration described in Section VIII B must be taken into account. We measure this by varying the response by $1\sigma$ within the statistical uncertainties of the fit and measure the difference in extracted top quark mass. We find this uncertainty to be 0.4 GeV/$c^2$.

The response calibration is derived from Monte Carlo experiments in which the expected number of $t\bar{t}$ events is calculated using the theoretical cross section as a function of $M_t$. The 10% uncertainty on the cross section is propagated to the final mass measurement, yielding an uncertainty related to the actual sample composition of 0.3 GeV/$c^2$.

Finally, we verified that the measurement is not sensitive to effects beyond leading order by measuring the mass in Monte Carlo experiments constructed with events with $M_t = 175$ GeV/$c^2$ generated by MC@NLO [43, 44] which includes next-to-leading order effects; the mean extracted mass in these experiments was $175.8 \pm 0.8$ GeV/$c^2$.

TABLE IV: Summary of systematic uncertainties.

| Source | $\delta M_t$ (GeV/$c^2$) |
|---|---|
| Jet Energy Scale | 2.6 |
| Background statistics | 1.2 |
| Background modeling | 0.8 |
| PDFs | 1.1 |
| FSR modeling | 0.5 |
| ISR modeling | 0.5 |
| Generator | 0.8 |
| Method | 0.4 |
| Sample composition uncertainty | 0.3 |
| Total | 3.4 |

### F. Summary of Systematic Uncertainties

The systematic uncertainties measured in the previous sections are summarized in Table IV. The total systematic uncertainty is 3.4 GeV/$c^2$ of which the single largest source is the jet energy scale, contributing 2.6 GeV/$c^2$.

## X. MEASUREMENT

The data sample in $\int Ldt = 340$ pb$^{-1}$ contains 33 candidate dilepton events. These candidates have individual likelihoods as seen in Figure 12. From the joint probability, we extract the uncorrected, unscaled mass,

$$\hat{M}_t = 166.4 \pm 3.4(\text{stat.})\ \text{GeV}/c^2.$$

After applying corrections to the central value and statistical uncertainty as derived in Section VIIID, the final result is

$$M_t = 165.2 \pm 6.1(\text{stat.})\ \text{GeV}/c^2.$$

The corrected joint probability curve can be seen in Figure 13. In Monte Carlo experiments where $M_t = 165$ GeV/$c^2$, 17% of the uncertainties are smaller than this value, see Figure 14.

## XI. CONCLUSION

We report the first application of a matrix-element based method to the measurement of the top quark mass in $t\bar{t}$ events containing two leptons. We measure

$$M_t = 165.2 \pm 6.1(\text{stat.}) \pm 3.4(\text{syst.})\ \text{GeV}/c^2,$$

which is the most precise determination to date of the top quark mass in dilepton events. This result is consistent with recent measurements of the mass in this channel at CDF using template methods,

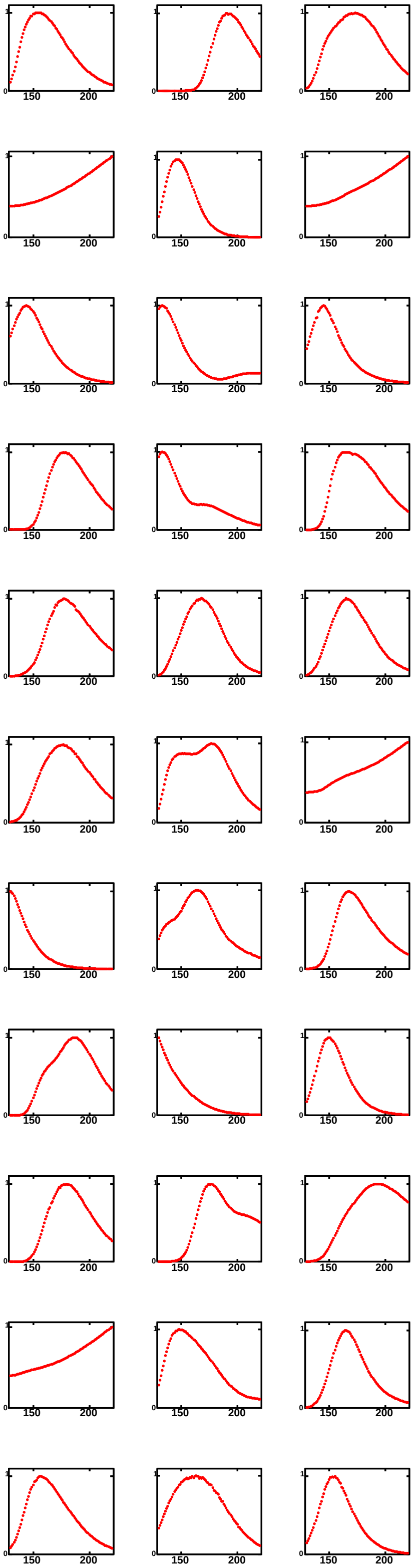

FIG. 12: Likelihood in top quark mass for the 33 candidate events in 340 pb$^{-1}$ of run II CDF data. Horizontal axis is top quark mass over the range of $M_t = 130$ to 220 GeV/$c^2$. Vertical axis is shown on a linear scale between 0 and 1.

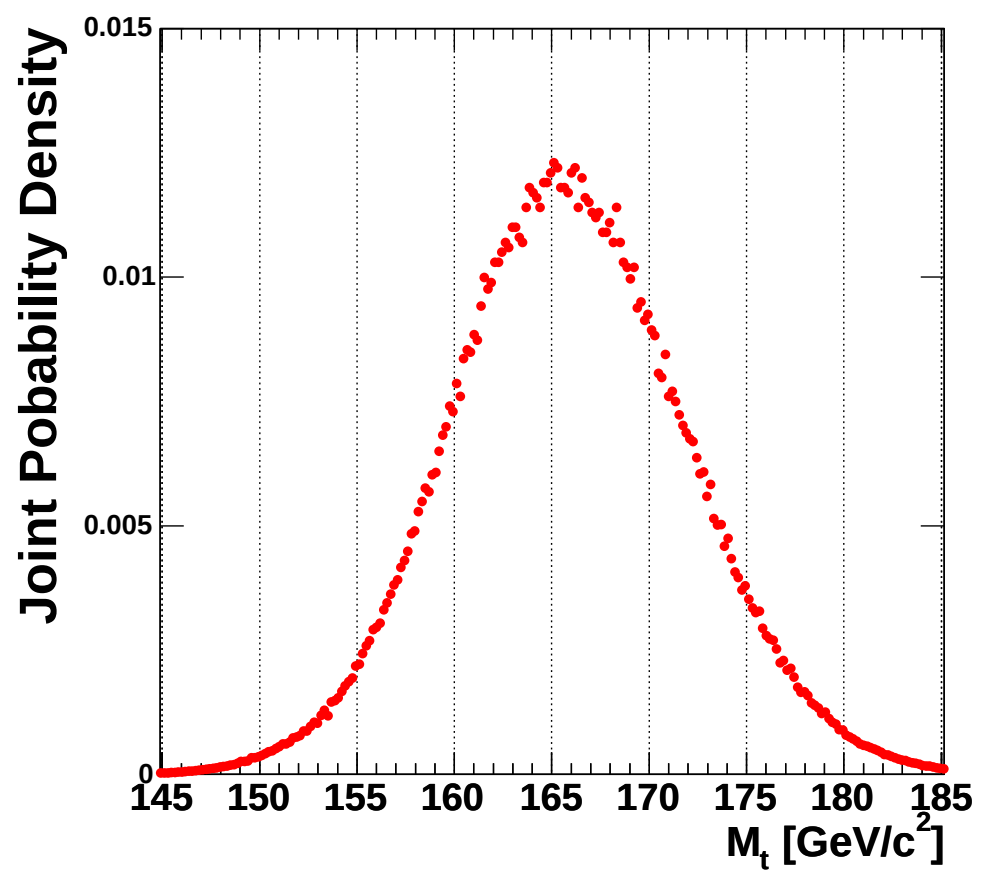

FIG. 13: Joint probability density for the 33 event data sample in 340 pb$^{-1}$ of run II CDF data as a function of the top quark mass. The bias and error corrections have been applied.

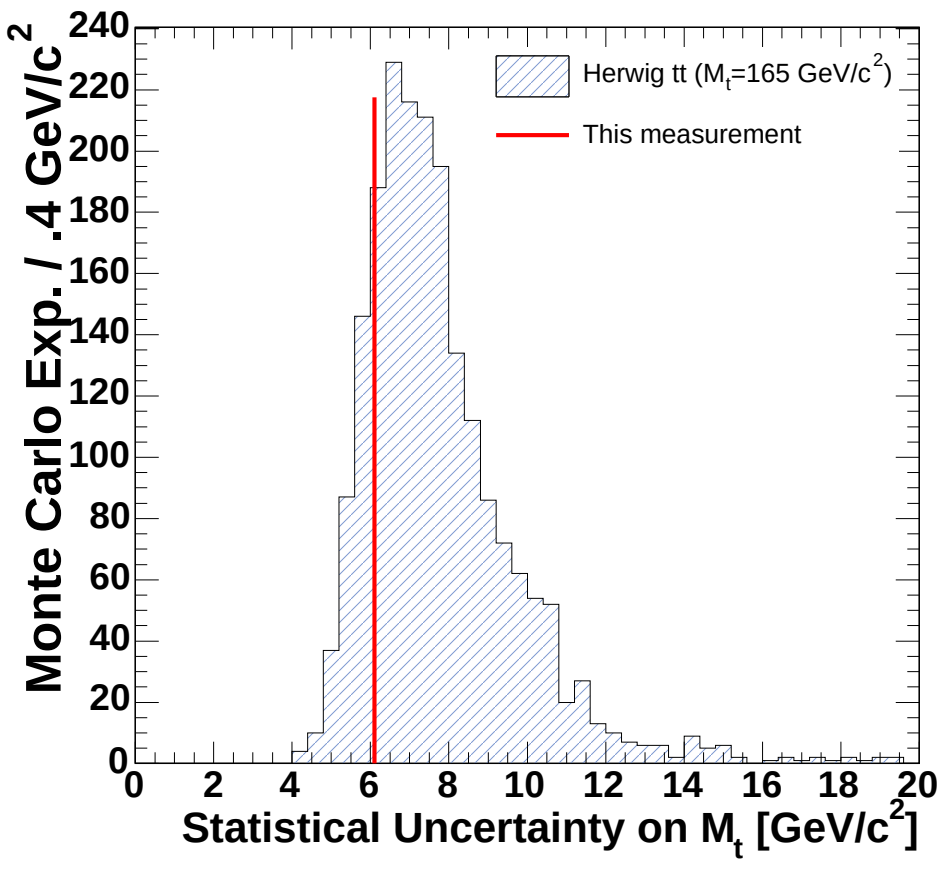

FIG. 14: Distribution of expected uncertainties for $M_t = 165$ GeV/$c^2$ in Monte Carlo experiments. The measured uncertainty is shown as the line; 17% of Monte Carlo experiments yield a smaller uncertainty.

$$M_t = 170.1 \pm 6.0(\text{stat.}) \pm 4.1(\text{syst.})\ \text{GeV}/c^2\ [45],$$

with measurements in run I from CDF,

$$M_t = 167.4 \pm 10.3(\text{stat.}) \pm 4.8(\text{syst.})\ \text{GeV}/c^2\ [6],$$

and DØ,

$$M_t = 168.4 \pm 12.3(\text{stat.}) \pm 3.6(\text{syst.})\ \text{GeV}/c^2\ [7].$$

This measured value is smaller than the current precision measurements of the mass in single lepton events,

$$M_t = 173.5^{+3.7}_{-3.6}(\text{stat.}) \pm 1.3(\text{syst.})\ \text{GeV}/c^2\ [46].$$

A global combination of the most precise measurements [47], however, suggests that current discrepancies are consistent with statistical fluctuations.

Extrapolating the use of the method to a future top quark dilepton data sample from the Tevatron corresponding to an integrated luminosity of $\int Ldt = 4$ fb$^{-1}$, the expected statistical uncertainty of this technique in Monte Carlo experiments is 2.5 GeV/$c^2$, for $M_t = 178$ GeV/$c^2$. In this regime, uncertainty in the jet energy scale would be the dominant source of uncertainty.

## Acknowledgments

We thank the Fermilab staff and the technical staffs of the participating institutions for their vital contributions. This work was supported by the U.S. Department of Energy and National Science Foundation; the Italian Istituto Nazionale di Fisica Nucleare; the Ministry of Education, Culture, Sports, Science and Technology of Japan; the Natural Sciences and Engineering Research Council of Canada; the National Science Council of the Republic of China; the Swiss National Science Foundation; the A.P. Sloan Foundation; the Bundesministerium für Bildung und Forschung, Germany; the Korean Science and Engineering Foundation and the Korean Research Foundation; the Particle Physics and Astronomy Research Council and the Royal Society, UK; the Russian Foundation for Basic Research; the Comisión Interministerial de Ciencia y Tecnología, Spain; in part by the European Community's Human Potential Programme under contract HPRN-CT-2002-00292; and the Academy of Finland.